\documentclass[aps,twocolumn,prd,10pt,nofootinbib,superscriptaddress]{revtex4-1}
\usepackage{aas_macros}
\usepackage{epsfig}
\usepackage{multirow}
\usepackage{bm}
\usepackage{latexsym}
\usepackage{natbib}
\usepackage{url}
\usepackage{balance}
\usepackage[T1]{fontenc}
\usepackage{lmodern}
\usepackage{color}
\usepackage{amsfonts,amssymb,amsmath}
\usepackage{graphicx,epsfig}
\usepackage{psfrag}
\usepackage{float}
\usepackage{subfigure}
\usepackage{caption}
\usepackage{subcaption}
\usepackage[citecolor = db]{hyperref}
\hypersetup{colorlinks=true}
\usepackage{mathtools}
\usepackage{enumitem}
\usepackage{float}
\usepackage{mathrsfs}
\usepackage[dvipsnames]{xcolor}
\usepackage[utf8]{inputenc}
\usepackage{comment}
\usepackage{cleveref}
 \crefname{section}{Section}{Sections}
 \crefname{equation}{Eq.}{Eqs.} 
 \crefname{figure}{Fig.}{Figs.}
 \crefname{appendix}{Appendix}{Appendices}
 \usepackage{soul}
\allowdisplaybreaks
\definecolor{db}{rgb}{0.0, 0.0, 0.62}
\definecolor{dm}{rgb}{0.7, 0.01, 0.7}
\definecolor{dr}{rgb}{0.55, 0.0, 0.0}

\begin{document}
\title{Effect of generic dark matter halo on transonic accretion onto galactic black holes}
\author{Avijit Chowdhury}
\email{avijit.chowdhury@rnd.iitg.ac.in}
\author{Gargi Sen}
\email{g.sen@iitg.ac.in}
\author{Sayan Chakrabarti}
\email{sayan.chakrabarti@iitg.ac.in}
\author{Santabrata Das}
\email{sbdas@iitg.ac.in}
\affiliation{Department of Physics, Indian Institute of Technology Guwahati, Assam-781039, India}
\begin{abstract}
The environment surrounding a black hole or black hole binaries is generally expected to play an important role in understanding various astrophysical phenomena around them. In this paper, we study relativistic, low angular momentum, inviscid, and advective hot accretion flow onto a galactic supermassive black hole dressed with a cold dark matter halo. Focusing on different relativistic dark matter distributions with an inner density spike, we analyze the effect of the dark matter halo on the topology and properties of the accretion flow. Our results show enhancement of disk luminosity in the presence of dark matter, which depends on the nature and properties (halo mass and compactness) of the dark matter distribution. Under the assumptions of our accretion model, the dominant contribution to the disk luminosity for compact and massive halos arises from the inner regions of the accretion flow. Consequently, our analysis indicates that luminosity measurements can serve as an effective probe of the underlying dark matter density spike.
\end{abstract}
\maketitle
\begin{balance}
\section{Introduction} \label{sec:Intro}
The stellar velocity dispersion of S stars (particularly the S2 star) around the centre of our Milky Way (Sgr.~A)~\cite{Genzel2010, 2008ApJ...689.1044G} together with the findings of the Event Horizon Telescope (EHT) Collaboration~\cite{EventHorizonTelescope:2022wkp} have established the existence of supermassive black hole (SMBH), Sgr.A* ($M_{\rm BH} \sim 4\times 10^6 M_\odot$) in the galactic centre with almost absolute certainty. Infact, almost all galaxies are believed to host a SMBH ($M_{\rm BH} \sim 10^6 - 10^{10} M\odot$)~\cite{Richstone:1998ky,Kormendy:2013dxa}, which plays a crucial role in their evolution~\cite{2017FrASS...4...42M}. Despite the overwhelming evidence of their existence, the formation mechanism and growth rate of these SMBHs remain an open question~\cite {Volonteri:2010wz}. The current theoretical models on the formation of SMBHs fail to explain the distribution of the black hole (BH) mass with distance as most of the SMBHs are found at high redshifts~\cite{Kormendy:2013dxa}. Although galaxy mergers in the early universe may lead to a possible explanation for the rapid growth rate of the SMBHs, it requires the presence of dark matter (DM)~\cite{Volonteri:2005fj} (see \cite{Sanders:2007mg, McGAugh:2024yiz, Skordis:2020eui} for alternatives).

The flattening of the rotation curves of spiral and disk galaxies at large radii provides strong evidence of the existence of DM~\cite{1940ApJ91273O}. Studies on the rotation curves of Sgr.~A~\cite{2020MNRAS.494.4291C, deSalas:2019pee} and other galaxies~\cite{Rubin:1980zd,Adams:2014bda} suggest that over $90 \%$ of the galactic mass is in the form of DM. Furthermore, CMB (Cosmic Microwave Background) observations indicate that DM constitutes over $85\%$ of the total matter density of the universe~\cite{Planck:2018vyg}. Though the exact nature and origin of DM remain a mystery, its abundance, coldness and lack of strong (non-gravitational) interaction with Standard Model particles are indisputable (see \cite{Lisanti:2016jxe} for a review). 
The cold dark matter (CDM) at the galactic centre gets redistributed by the SMBH, whose strong gravity results in a significant increase ({\it spike}) in the local mass density of the CDM outside the event horizon. Assuming the central SMBH (in the absence of DM) to be represented by the Schwarzschild geometry, a fully general relativistic analysis suggests that the DM density distribution must vanish at twice the Schwarzschild radius~\cite{Sadeghian:2013laa} (see also \cite{Gondolo:1999ef} for a Newtonian analysis). Thus, the galactic cores serve as a unique laboratory to probe the effect of DM on strong gravity phenomenology~\cite{deLaurentis:2022oqa}. 

At low redshift, SMBHs primarily grow through the process of accretion, where (luminous) matter from their surrounding environment falls onto them, gradually increasing their mass. These are observed as active galactic nuclei (AGN). The accretion rate influences AGN properties, launching outflows that may shape the host galaxy's growth~\cite{Bower:2005vb, 2017FrASS...4...42M, 2017NatAs...1E.165H}. The environmental dependence of AGN activity can provide powerful constraints on evolutionary scenarios of SMBH fueling and feedback~\cite{Powell:2022tpv}. The present work is an attempt in this direction to qualitatively analyze the influence of cold collisionless DM halo, particularly the DM over-density/spike on the {\it advection dominated transonic hot accretion flow} of (luminous) matter onto the central SMBH. 

 Accretion of luminous matter onto BHs and other compact objects is believed to be one of the primary mechanisms to power astrophysical sources, such as  X-ray binaries~\cite{Vidal-etal1973} and AGNs~\cite{Peterson-1997, Fabian-1999}. The accretion flow starts subsonically far away from the BH (outer edge of the accretion disk) and spirals inward, approaching the speed of light as it reaches the event horizon. Thus, the accretion flow is transonic in nature~\cite{Novikov-Thorne1973, Matsumoto-etal1984, Fukue-1987, Chakrabarti-1989,  Narayan-etal1996}. 
Ichimaru \cite{ichimaru} first pointed out that the energy generated by viscous dissipation will increase the temperature of the accretion flow rather than being radiated away. As a result, the advection-dominated accretion flow (ADAF) becomes hot. This hot accretion must also be thermally stable. The dynamical and radiative characteristics have been extensively investigated~\cite{Narayan-Yi1994, Narayan-Yi1995a, Narayan-Yi1995b, Abramowicz-etal1995}. Though the hot accretion model is usually considered to be radiatively inefficient, with their cooling time being longer than the infall time, it can successfully describe the hard and quiescent states of the supermassive black hole in the galactic centre, low-luminosity AGNs and black hole  X-ray binaries. On the other hand, in thin disk or cold disk model, the gas temperature is lower than the virial temperature \cite{Shakura-Sunyaev1973, Novikov-Thorne1973, Lynden-Bell-Pringlel1974, Pringle-1981, Frank-etal2002, Kato-etal2008, Abramowicz-2013, Blaes-2014}, making the disk geometrically thin and optically thick resulting in thermal blackbody-like radiation.

Since the dynamics of accretion flow around compact objects is strongly governed by the geometry of the surrounding spacetime, one must accurately model the modification of the spacetime geometry due to the relativistic distribution of the DM particles around the SMBH at the galactic centre. The first step in this scheme is a good description of the galactic DM density profile, guided by observations and large-scale simulations~\cite{doi:10.1142/4886}. Assuming a Hernquist (HQ) type DM density distribution~\cite{Hernquist:1990be}, Cardoso et al. ~\cite{Cardoso:2021wlq} proposed an exact analytical solution representing SMBH minimally coupled to the anisotropic dark matter fluid using the Einstein Cluster model~\cite{Geralico:2012jt,Einstein:1939ms} within general relativity (GR). The spacetime solution obtained in ~\cite{Cardoso:2021wlq} was used in~\cite{Cardoso:2022whc} to develop a generic, relativistic formalism to study gravitational-wave emission by extreme-mass-ratio inspirals. Going beyond the Hernquist class of DM distribution, a fully numeric pipeline was developed in ~\cite{Figueiredo:2023gas, Speeney:2024mas} to treat generic density perturbation around spherically symmetric galactic SMBH.  See Refs.~\cite{Retana-Montenegro:2012dbd, Jusufi:2020cpn, Igata:2022rcm, Dai:2023cft, Xavier:2023exm, Myung:2024tkz, Kazempour:2024lcx, Chen:2024lpd, Tan:2024hzw, Zhao:2024bpp, Mollicone:2024lxy, Amancio:2024mrt, Stuchlik:2021gwg, Rahman:2023sof,Faraji:2024ein,Macedo:2024qky,Liu:2022lrg,Singha:2023lum,Patra:2025mnj,Ranjbar:2025yml,Ranjbar:2025esx} for studies on the influence of DM halos on different spacetime properties and phenomenology. In the present work, we follow the numerical framework developed in~\cite{Figueiredo:2023gas} to model the spacetime around a static spherically symmetric SMBH, dressed with a DM halo, with different relativistic density distributions. 

Our results indicate that the change in geometry due to the feedback from the relativistic distribution of DM halo causes a significant change in the effective gravitational potential, leading to deviation in the topology and properties of the accretion flow. This results in substantial differences in the spectral energy density and the bolometric luminosity, particularly for highly compact halos with large halo mass. The quantitative deviation of the flow properties and luminosity from that of the vacuum Schwarzschild BH depends on the exact nature of the halo density profile (and the DM spike model).
In the present work, we consider three different DM density profiles, namely, Hernquist, NFW, and Einasto, along with a fully relativistic model of the DM spike based on Hernquist-type distribution.

The paper is arranged as follows. In \cref{sec: BG-density}, we discuss the geometry of the spacetime around the central SMBH with different numerically and observationally supported Halo density profiles. In \cref{sec:assume}, we highlight the assumptions involved in the analysis and discuss the governing equation for transonic accretion flow. \cref{sec:crit-sol} deals with the occurrence of the sonic points and the topology of the global accretion solution. Subsequently, in \cref{sec:param,sec:solprop,sec:lumi}, we discuss the results concerning the parameter space, solution properties and the luminosity of the global accretion solution, respectively. Finally, we conclude with a brief summary and discussion of our results in \cref{sec:concl} highlighting their importance and prospects. 

Throughout the paper, we use geometric units ($G=c=1$) unless stated otherwise.
\section{Background geometry and density profiles}\label{sec: BG-density}
This section briefly reviews the formalism to construct an asymptotically flat, static, spherically symmetric BH solution embedded in a DM halo with different density distributions.
\subsection{Background Geometry}\label{BG}
We model the spacetime geometry of a BH by the line element,
\begin{equation}\label{eq:metric}
			ds^2 =  -f(r) dt^2 + \frac{dr^2}{1-\frac{2m(r)}{r}}+r^2d\Omega^2~,
\end{equation}
where $d\Omega^2$ represents the metric on a unit two-sphere. 

Following~\cite{Cardoso:2021wlq, Figueiredo:2023gas, Speeney:2024mas, Chakraborty:2024gcr, Pezzella:2024tkf}, we employ a generalized Einstein cluster formalism~\cite{Einstein:1939ms, Geralico:2012jt}. 
Einstein cluster is a collection of collisionless particles in all possible circular geodesics.
The average stress energy tensor is given by, $$\left< T^{\mu\nu}\right>=\frac{n}{m_p}\left<P^{\mu} P^{\nu}\right>~,$$ where $n$ is the proper number density of particles with rest mass $m_p$ and $P^\mu$ is the four-momentum satisfying the geodesic equation. The averaging $⟨\ldots⟩$ is done over all trajectories with all the directions and phases passing through a spatial point where the energy-momentum
tensor of the orbiting particles is computed. This averaging ensures that the system remains static and spherically symmetric. The number density, $n(r)$, on a given shell is independent of that in the other shells. This construction
is equivalent to an anisotropic material with only tangential pressure $P_t$ and zero radial pressure. Thus, we assume the metric in \cref{eq:metric} to be a solution of the Einstein's equations,
\begin{equation}\label{eq:EFE}
	G_{\mu\nu}=8\pi T^\text {env}_{\mu\nu},
\end{equation}
where
\begin{equation}\label{eq:Tmunu}
\left({T}^\mu_{\nu}\right)^{\rm env}= {\rm diag}(-\rho_{{\rm DM}} (r),0,P_t(r), P_t(r))~,
\end{equation}
is the anisotropic stress-energy tensor encoding the properties of the environment in terms of the density $\rho_{{\rm DM}}(r)$ and the tangential pressure $P_t(r)$ of the matter distribution. The zero radial pressure of the distribution is also indicative of the negligible DM accretion rate within the timescale of baryonic accretion. For a given density profile, the continuity equation determines the mass profile as 
\begin{equation}\label{eq:mprime}
  m'(r)=4 \pi r^2 \rho_{\rm DM}(r)~,  
\end{equation}
whereas the metric function $f(r) $ and the tangential pressure $P_t (r)$ are determined by the $rr$ component of the field equations  (\cref{eq:EFE}) and the Bianchi identities, respectively,
\begin{equation}   \label{eq:fprime}
    \frac{f'(r)}{f(r)}= \frac{2 m(r)/r}{r-2 m(r)}~,
\end{equation}
\begin{equation}\label{eq:pt}
    P_t(r)=\frac{m(r)/2}{r-2 m(r)}\rho_{\rm DM}(r)~.
\end{equation}
The metric functions $f(r)$ and $m(r)$ completely specify the geodesic structure of the spacetime. 

The  stationarity and spherical symmetry of the metric \eqref{eq:metric} implies the existence of a timelike and a spacelike Killing vector associated with the conserved quantities, the specific energy and the specific angular momentum at infinity.
The radius of the innermost stable circular orbit for massive particles is given by the roots of the equation,
\begin{equation}\label{eq:r-isco}
r^2m'(r)+r m(r) -6 m^2(r)=0~,
\end{equation}
with the associated angular frequency,
\begin{equation}\label{eq:omega-isco}
\Omega_{\rm ISCO}= \left[\frac{f(r) m(r)}{r^2(r-2m(r))}\right]_{r=r_{\rm ISCO}}~.
\end{equation}
The radius of the  lightring (unstable null circular geodesic) is determined by the roots of the equation, $r=3 m(r)$, with the associated angular frequency
\begin{equation}\label{eq:omega-LR}
    \Omega_{\rm LR}=\frac{\sqrt{f(r_{\rm LR})}}{r_{\rm LR}}~.
\end{equation}

To evaluate the metric functions $f(r)$ and $m(r)$, we follow the numerical procedure outlined in~\cite{Speeney:2024mas} motivated by the solutions obtained in ~\cite{Konoplya:2022hbl, Shen:2023erj, Maeda:2024tsg}. We start with a specific choice of the DM density profile $\rho_{\rm DM}(r)$ (as outlined in \cref{sec:density-profile}) and 
numerically integrate \cref{eq:mprime} from $r=2M_{\rm BH}$ to $r_{\infty}=10^{16} M_{\rm BH}$ (corresponding to our numerical infinity) to determine $m(r)$. We then use $m(r)$ in \cref{eq:fprime} to evaluate $f(r)$ by integrating backwards from $r_\infty$ with the boundary conditions,
\begin{eqnarray}
    m_{r_\infty}=M_{\rm BH}+ M_{\rm halo}~,\\
    f(r_\infty)= 1-\frac{2m(r_\infty)}{r}~,
\end{eqnarray}
where $M_{\rm halo}$ is the total mass of the DM environment surrounding the BH. Note that by construction, the event horizon is $r_{\rm h}= 2 M_{\rm BH}$.
 The obtained metric functions can then be used in \cref{eq:pt} to determine the tangential pressure. Similarly, the location and angular momentum at the ISCO and lightring can also be determined explicitly, using \cref{eq:r-isco,eq:omega-isco,eq:omega-LR} respectively.


\subsection{Environmental density profiles} \label{sec:density-profile}

Since we are interested in studying the effect of collisionless CDM distribution on the accretion onto the central BH, we consider two of the most widely studied CDM density profiles, the Hernquist~\cite{Hernquist:1990be} and the Navarro-Frenk-White (NFW)~\cite{Navarro:1994hi,Navarro:1996gj} distributions. Whereas the Hernquist profile is mostly used to model the S\'ersic profile observed in bulges and elliptic galaxies, the NFW profile is mainly used for galaxies with the largest content of DM~\cite{Kormendy:2013dxa}.
Both these profiles can be parametrically described as ~\cite{Taylor:2002zd},
\begin{equation}\label{eq:density-prof}
    \rho_{{\rm DM}}(r)=\rho_{{\rm DM}}^0 (r/a_0)^{-\gamma}\left[1+\left(r/a_0\right)^{\alpha}\right]^{(\gamma-\beta)/\alpha},
\end{equation}
where $\rho_{{\rm DM}}^0=2^{(\beta-\gamma)/\alpha} \rho_{{\rm DM}}(a_0)$ is a scale factor with $a_0$ being the scale radius of the halo.
The parameters $\beta$ and $\gamma$ respectively govern the dependence of the profile at large and small radii, whereas $\alpha$ determines the sharpness of the change of the profile slope at $a_0$~\cite{Taylor:2002zd}.
The Hernquist and NFW density profiles correspond to the parameter set $(\alpha,\beta,\gamma)$ = $(1,4,1)$ and  $(\alpha,\beta,\gamma)$ = $(1,3,1)$, respectively. Since the total mass of the NFW profile is logarithmically divergent, we use a cutoff radius $r_c$ for the dark matter distribution, such that $M_{\rm halo}(r>r_c)=0$. Henceforth, for the NFW density profile, we will use two values of the cutoff radius: (a) $r_c=5 a_0$ (referred to as NFW), (b)$r_c=a_0$ (referred to as NFW1) . We also use a third-density profile named, Einasto profile~\cite{1969Afz.....5..137E,Acharyya:2023rnq}, that, as per recent $N$-body CDM simulations, provides an excellent fit to a wide range of dark matter halos. The Einasto profile is given by,
\begin{equation}\label{eq:einasto}
    \rho_{{\rm DM}}(r)=\rho_{{\rm DM}}^e {\rm exp}\left\{-d_n \left[(r/r_e)^{1/n}-1\right]\right\}~,
\end{equation}
 where $r_e$ denotes the radius of the sphere containing half of the total mass, and $\rho_{{\rm DM}}^e$ is the mass density at $r= r_e$ with $n=6$ and $d_n=53/3$~\cite{Graham:2005xx, Prada:2005mx}. Throughout the work, we set $r_e=a_0$ for brevity.

Due to the adiabatic accretion growth of a seed nonrotating BH sitting at the core, the DM distribution is expected to develop an overdensity with a sharp cutoff close to the horizon at $r=4M_{\rm BH}$ \cite{Sadeghian:2013laa,Gondolo:1999ef}. Following ~\cite{Speeney:2024mas}, we model the overdensity in two different ways:
\begin{itemize}
    \item[1.] We multiply the DM density profile with the cut-off factor $(1-4M_{\rm BH}/r)$.\footnote{The radius $r=4M_{\rm BH}$ corresponds to the radius of the unstable circular orbit of a marginally bound particle with angular momentum per unit mass $L=4M_{\rm BH}$. Any particle with energy $\mathcal{E}\leq 1$ and angular momentum $L\geq 4 M_{\rm BH}$ has an inner turning point at $r\geq 4M_{\rm BH}$. So, a particle reaching $r=4M_{\rm BH}$ with $\mathcal{E}=1$ is necessarily captured by the BH.} The choice of the cutoff factor ensures that the dominant energy condition is satisfied~\cite{Shen:2024qbb,Datta:2023zmd,Speeney:2024mas} in the region, $2M_{\rm BH}<r<4M_{\rm BH}$. 
    \item[2.] We consider a fully relativistic DM spike model introduced in~\cite{Speeney:2022ryg}. We concentrate on the Hernquist subclass of the DM spike i.e., we consider the adiabatic growth of the DM spike starting from a Hernquist distribution given by~\cite{Speeney:2024mas},
    \begin{equation}
    \begin{split}
    \label{eq:HQSpk}
        \rho_{{\rm DM}}(r)=&\bar{\rho}_{{\rm DM}}(r)\frac{M_{{\rm halo}}}{M_{{\rm BH}}^2}\\
        &\times\left( \int_{4 M_{{\rm BH}}}^{r_c} 4 \pi r^2 \bar{\rho}_{{\rm DM}}(r) dr\right)^{-1},
        \end{split}
    \end{equation}
    with
    \begin{align}\label{eq:eqspk1}
        \bar{\rho}_{{\rm DM}}(r)=&\left(1-\frac{4M_{{\rm BH}}}{r}\right)^\alpha\left(\frac{r M_{{\rm BH}}}{M_{{\rm halo}} a_0}\right)^\beta  \nonumber\\
        &\times\left(1+\frac{r M_{{\rm BH}}}{M_{{\rm halo}} a_0}\right)^\gamma,
    \end{align}
    where $\alpha=2.366$, $\beta=-2.320$ and $\gamma=-1.370$. The functional dependence of $\bar\rho_{{\rm DM}}(r)$ on $a_0$ and $M_{\rm halo}$ was determined empirically by studying the behaviour of an ensemble of DM profiles computed numerically for selected values of $M_{\rm halo}$ and $a_0$~\cite{Sadeghian:2013laa}. The numerical values of $\alpha$, $\beta$ and $\gamma$ were then determined by fitting with a numerical spike with $M_{\rm halo}=10^4 M_{{\rm BH}}$ and $M_{\rm halo}/a_0=0.001$.
\end{itemize}

Fig.~\ref{fig:density-prof} shows the representative plots of different DM distributions for different values of the halo mass and scale radius. We note that the Hernquist and the NFW profiles behave almost similarly, with the Hernquist density being greater than the NFW density at a fixed radius. The NFW distribution terminated at $a_0$ (NFW1) has a higher density than the one terminated at $5a_0$ since the total halo mass remains constant. For the halo-compactness ($M_{\rm halo}/a_0$) considered in \cref{fig:density-prof}, the Hernquist type spike profile has the highest density of DM close to the BH, however for more compact halos, the situation is reversed \cite{Speeney:2024mas}. 

In the subsequent sections, we study the accretion flow of the baryonic matter onto the galactic BH dressed with a dark matter halo. We again stress that the DM influences the accretion flow of the baryonic matter by modifying spacetime geometry. 

Since there are multiple length scales involved in the problem, we maintain a strict separation of the length scales, $M_{\rm BH}\leq M_{{\rm halo}}\leq a_0$ and $r_{\rm edge}\leq a_0$, where $r_{\rm edge}$ is the outer edge of the accretion disk. Note that in all the cases, we treat $M_{\rm halo}$ and $a_0$ as free model parameters for a qualitative analysis.

\begin{figure*}[htbp!]
    \includegraphics[width=\linewidth]{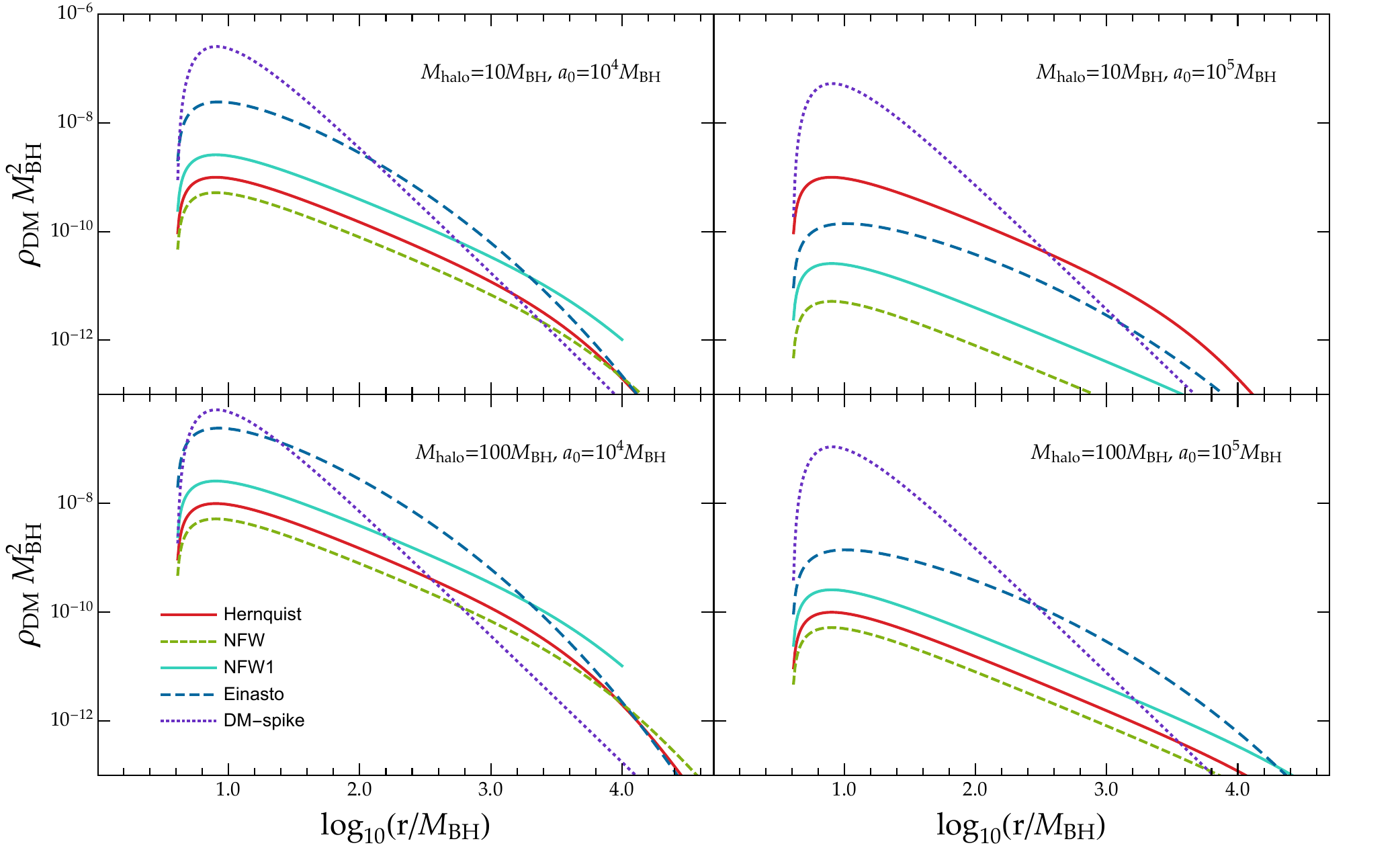}
    \caption{Halo density profiles for different values of the $M_{\rm halo}$ and $a_0$. For the NFW profile, two values of the truncation radius have been used,  $r_c=5a_0$ (NFW) and $r_c=a_0$ (NFW1). The DM spike profile is of the Hernquist subclass. See the text for details.}
    \label{fig:density-prof}
\end{figure*}

\section{Assumptions and Model Equations}\label{sec:assume}

In this section, we present the basic equations governing the \textit{advection dominated, steady, inviscid, hot accretion flow of non-magnetic and optically thin}  plasma~\cite{,Yuan:2014gma} onto the central BH. The spherical symmetry of the background geometry (\cref{eq:metric}) allows us to consider a planar accretion flow (for brevity we choose to be the equatorial plane). The energy-momentum tensor and the four-current of the fluid is given by,
\begin{equation}\label{eq:Tmunu}
    T^{\mu\nu}= (e+p)u^\mu u^\nu+p g^{\mu\nu} \quad \mbox{and} \quad j^\mu =\rho u^\mu
\end{equation}
where $e$, $p$, $\rho$, and $u^\mu$ are the local energy density, pressure, mass density and four-velocity of the fluid element, respectively, with $u^\mu u_\mu=-1$.
The conservation of the energy-momentum tensor and the mass-flux provide the relativistic hydrodynamical equations that completely determine the accretion flow. These are given by,
\begin{equation}
T^{\mu\nu}_{;\nu}=0, \quad \mbox{and}\quad j^{\nu}_{;\nu}=0.
\end{equation}
Using the projection tensor, $h_{\mu\nu}=g_{\mu\nu}+u_{\mu}u_{\nu}$, projecting the conservation equation onto the hypersurface orthogonal to the flow velocity, we get the relativistic Euler equation as,
\begin{equation}\label{eq:euler}
    h^\alpha_\mu T^{\mu\nu}_{;\nu}=(e+p)u^{\nu}u^{\alpha}_{;\nu}+\left(g^{\alpha \nu}+u^{\alpha}u^{\nu}\right)p_{,\nu}=0.
\end{equation}
Similarly, the component of the conservation equation along the flow velocity yields the energy equation,
\begin{equation}\label{eq:1stlaw}
    u_\mu T^{\mu\nu}_{;\nu}=u^\mu\left[\left(\frac{e+p}{\rho}\right)\rho_{;\mu}+ e_{;\mu}\right]=0.
\end{equation} 
Since the temperature of the accretion flow close to the BH is of the order $\sim 10^{10}-10^{11}{\rm K}$, the accretion flow is thermally relativistic. Thus, we use the relativistic equation of state (REoS) ~\cite{Chattopadhyay:2008xd} with variable adiabatic index to relate the density and pressure of the flow,
\begin{equation}\label{eq:REoS}
    e= \frac{\rho}{\tau}\tilde{f}, \quad p=\frac{2 \rho \Theta}{\tau},
\end{equation}
with, $\tau=1+m_i/m_e$ and 
\begin{align}\label{eq:f}
    \tilde{f}=&\left[1+\Theta\left(\frac{9\Theta +3}{3\Theta+2}\right)\right]\nonumber\\
    &+\left[\frac{m_i}{m_e}+\Theta\left( \frac{9 \Theta m_e +3 m_i}{3 \Theta m_e +2 m_i}\right) \right],
    \end{align}
    where $m_e$ and $m_i$ are the masses of electron and ion, respectively, and $\Theta~(=k_B T/m_e c^2)$ is the dimensionless temperature. In the above equation and all subsequent calculations, we implicitly assume the number density of the electrons and ions to be the same. Using the REoS in \cref{eq:REoS}, we express the associated sound-speed as $C_s=\sqrt{2 \Gamma \Theta/ (\tilde{f}+2\Theta)}$, where $\Gamma~(=(1+N)/N)$ is the adiabatic index with $N=(1/2)(d\tilde{f}/d\Theta)$ being the polytropic index of the flow. Since the accretion flow is assumed to be confined in the equatorial plane $(\theta=\pi/2, u^\theta=0)$,
    we write the radial component of the relativistic Euler equation as, 
    \begin{equation}
        \label{eq:Euler-r}
        v\gamma_{v}^{2}\frac{dv}{dr}+\frac{1}{h\rho}\frac{dp}{dr}+\frac{d\Phi^{\textrm{eff}}_{e}}{dr}=0,
	\end{equation}
    where $\gamma_v=1/\sqrt{1-v^2}$ is the Lorentz factor associated with the fluid three velocities in the corotating frame, $v=\gamma_{\phi}v_{r}=\sqrt{u^r u_r/ \left[u^t u_t(\lambda^2 f(r)/r^2 -1)\right]}$. The quantity $h~(=(e+p)/\rho)$ is the specific enthalpy and  $\Phi^{\textrm{eff}}_{e}$ is the effective potential \cite{Dihingia:2018tlr} given by,
    \begin{equation}\label{eq:eff-pot}
        \Phi^{\textrm{eff}}_{e}=1-\frac{1}{2}\ln \left( \frac{1}{f(r)}-\frac{\lambda^2}{r^2} \right),
    \end{equation}  
    As evident from \cref{eq:eff-pot}, the effective potential depends on the spacetime geometry and the specific angular momentum of the flow. 
    
    The symmetry of the background spacetime allows us to consider a stationary and axisymmetric accretion flow. Thus, we define the specific energy ($\mathcal{E}$) and the conserved angular momentum ($\mathcal{L}$) associated with the timelike and spacelike Killing vectors ($\partial_t$) and ($\partial_\phi$) respectively as,
    \begin{equation}
        -h u_t=\mathcal{E}, \quad{\rm and} \quad h u_{\phi}=\mathcal{L}~.
    \end{equation}
    We also define the specific angular momentum as $\lambda=\mathcal{L}/\mathcal{E}$, which is conserved along the flow streamline.

    The effect of the DM density profile on the effective potential is shown in \cref{fig:eff-pot}.
    \begin{figure*}[htbp!]
            \includegraphics[width=\textwidth]{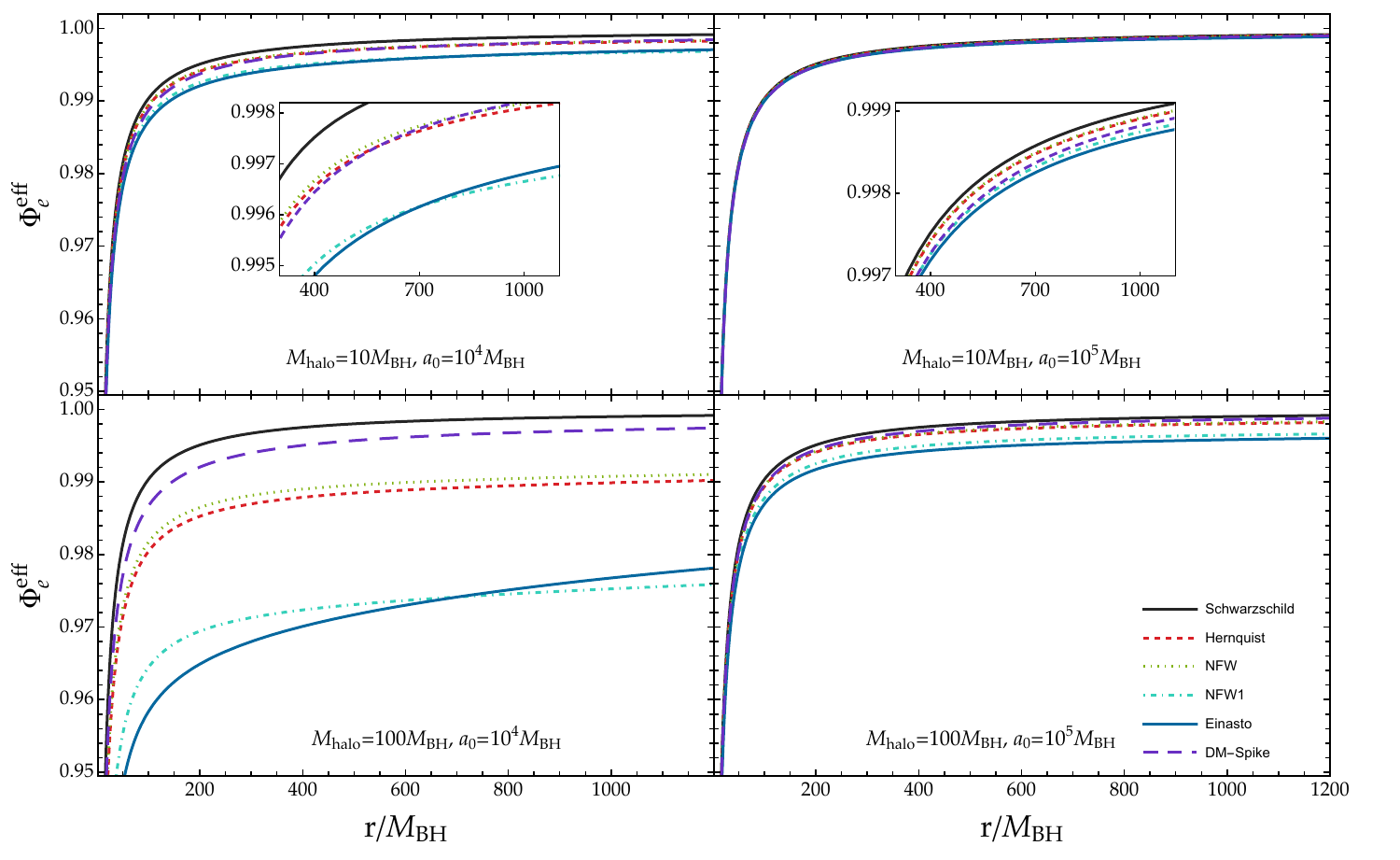}
        \caption{Variation of the effective potential ($\Phi^{\rm eff}_e$) as a function of the radial coordinate $r$ in the unit of the BH mass for different DM halo density profiles. Here, we choose $\lambda=3$. See the text for details.
        }
        \label{fig:eff-pot}
    \end{figure*}
We observe that compared to the vacuum Schwarzschild BH, the presence of the DM halo lowers the effective potential at any given radius. Thus, a flow with lower specific energy can overcome the potential barrier. For a fixed total halo mass as $a_0$ increases (DM distribution gets more dilute), $\Phi_e^{\rm eff}$  approaches the Schwarzschild limit for all DM distributions. Again, for a given $a_0$, as the halo mass increases, the deviation of the effective potential of each of the DM distributions from the Schwarzschild value increases. However, for a fixed compactness, ($\Psi=M_{\rm halo}/a_0$), the higher the halo mass, the lower the effective potential. Though the effective potential for the different DM distributions follows a similar trend, they differ largely among themselves. We observe that close to the central BH, the Einasto profile gives rise to the smallest effective potential. Far away from the central BH the effective potential for the NFW1 profile is usually the smallest. The effective potential for HQ-type DM spike distribution strongly depends on the halo mass as shown in \cref{app:DMspk}. 

The radial component of the entropy-generation equation (\cref{eq:1stlaw}) gives
\begin{equation}\label{eq:ent-gen}
    \left(\frac{e+p}{\rho}\right) \frac{d\rho}{dr}-\frac{d e}{dr}=0.
\end{equation}
The integrated form of the continuity equation (\cref{eq:Tmunu}) gives the conserved mass accretion rate of the flow as,
\begin{equation}\label{eq:Mdot}
    \dot{M}=4 \pi r \rho v \gamma_v H \sqrt{f(r)},
\end{equation}
where $H=\sqrt{p r \left(r^2-\lambda^2 f(r)\right)/\rho}$ is the local half-thickness of the disk.
Following~\cite{Kumar:2017flo, Chattopadhyay:2016kcz}, we define the entropy accretion rate of the flow as,
\begin{equation}
    \dot {\cal M}=\frac{\dot{M}}{4 \pi \mathcal{K}}= \frac{\rho}{\mathcal{K}}v \gamma_v H r \sqrt{f(r)},
\end{equation}
where the density is obtained by integrating \cref{eq:ent-gen} using \cref{eq:REoS,eq:f},
\begin{equation}
\begin{split}\label{eq:rho}
    \rho=&\mathcal{K} \exp\left( \frac{f-\tau}{2 \Theta}\right)\Theta^{3/2}\left(3 \Theta + 2\right)^{3/4} \\
    &\times \left( 3 \Theta+2 m_i/m_e\right)^{3/4}~,
    \end{split}
\end{equation}
$\mathcal{K}$ being the constant of integration.
Since we have considered an adiabatic accretion flow, the entropy accretion rate $\dot{\mathcal{M}}$ remains constant throughout the flow streamline. Furthermore, using \cref{eq:ent-gen,eq:Mdot}, we express the dimensionless temperature gradient as,
\begin{align}\label{eq:temp-grad}
    \frac{d \Theta}{dr}=&-\frac{2 \Theta}{2 N+1} \nonumber \\
    &\times\left[\frac{1}{r}\left(\frac{5}{2}+\frac{\lambda^2}{r^2-\lambda^2}\right)+\frac{\gamma_v^2}{v}\frac{d v}{dr}  +\frac{f'(r)}{2f(r)}  \right].
\end{align}
Using \cref{eq:Euler-r,eq:temp-grad}, we obtain the wind equation in terms of the sound speed as, 
\begin{equation}\label{eq:wind}
    \frac{dv}{dr}=\frac{\mathcal{N}}{\mathcal{D}},
\end{equation}
where,
	\begin{align}
		&\mathcal{N}=\frac{2 C^{2}_{s}}{\Gamma +1}\left[\frac{1}{r}\left(\frac{5}{2}+\frac{\lambda^2}{r^2-\lambda^2}\right)+\frac{f'(r)}{2f(r)}\right]-\frac{d\Phi^{{\rm eff}}_{e}}{dr},\label{eq:N}\\
		&\mathcal{D}=\gamma_{v}^{2}\left(v-\frac{2 C^{2}_{s}}{v(\Gamma +1)}\right)\label{eq:D}.
    \end{align}

\section{Critical point analysis and Global accretion solution}\label{sec:crit-sol}

The accretion flow starts subsonically from the outer edge of the accretion disk ($v<C_s$) and continues smoothly along a streamline with the infall velocity approaching the speed of light as it reaches the event horizon. Thus, during the course of accretion, the flow turns supersonic at a specific radius depending on the background geometry, the specific angular momentum and energy of the flow. Hence, the accretion flow is \textit{transonic} in nature. At the critical radius ($r_c$), both $\mathcal{D}$ and $\mathcal{N}$ in \cref{eq:wind} must vanish identically to maintain the smoothness of the accretion flow. 
Setting the critical point conditions $\mathcal{D}=\mathcal{N}=0$, we get,
\begin{align}\label{eq:crit-cond}
v_c^2=&\frac{2 C_{sc}^2}{\Gamma_c + 1},\\
     C_{sc}^2=&\frac{\Gamma_c+1}{2}\left(\frac{d\Phi_e^{\rm eff}}{dr}\right)_{c}  \nonumber\\
     &\times\left[\frac{1}{r_c}\left(\frac{5}{2}+\frac{\lambda^2}{r_c^2-\lambda^2}\right)+\frac{f'(r_c)}{2f(r_c)}\right]^{-1},
 \end{align}
where the subscript $c$ denotes the corresponding quantity being evaluated at the critical point $r=r_c$.
Since $dv/dr$ takes ``0/0'' form at the critical point, we use the L'H$\hat{\rm o}$spital's rule to calculate the radial velocity gradient of the flow, $\left( dv/dr\right)_{r_c}$. At a critical point, if both values of $\left(\frac{dv}{dr}\right)_{r_c}$ are real and of opposite sign, the critical point is referred to as \textit{saddle} type, whereas if they are of the same sign, the critical point is referred to as \textit{nodal} type. In case, the radial velocity gradient at a critical point becomes imaginary, the corresponding critical point is referred to as \textit{spiral} type. Henceforth, we only consider saddle-type critical points since they are stable against small-amplitude perturbations~\cite{10.1093/mnras/260.2.317}. If a saddle-type critical point develops close to the event horizon, it is referred to as the inner critical point ($r_{\rm in}$), whereas if it develops away from the event horizon, it is referred to as the outer critical point ($r_{\rm out}$).

Using the flow velocity ($v_c$) and temperature ($\Theta_{\rm c}$) evaluated at the critical point as initial conditions, we integrate \cref{eq:temp-grad,eq:wind} from the critical point to the outer edge ($r_{\rm edge}$) and again from the critical point to the horizon ($r_{\rm h}$). Finally, we combine both these segments to get the global transonic accretion solution.
 If the accretion flow contains only an inner critical point ($r_{\rm in}$), the corresponding flow topology is referred to as I-type, whereas if the flow contains only an outer critical point ($r_{\rm out}$), it is called O-type. 
Depending on the background geometry and flow parameters, the accretion flow may as well contain both the inner and outer critical points simultaneously. In such cases, if the entropy accretion rate at $r_{\rm in}$ is greater than that at $r_{\rm out}$ (${\dot {\cal M}}_{\rm in} > {\dot {\cal M}}_{\rm out}$), the flow is referred to as A-type, whereas for the reversed case, when ${\dot {\cal M}}_{\rm in} < {\dot {\cal M}}_{\rm out}$, the flow is called W-type \cite{Chakrabarti-1989,Chakrabarti-1996}. We represent all these types of solutions in \cref{sec:soln-topo}. For an A-type solution, the flow coming from the outer edge ($r_{\rm edge}$), connects the horizon ($r_{\rm h}$) only through the outer critical point ($r_{\rm out}$). On the other hand, for a W-type solution, the accretion flow passing through $r_{\rm in}$ only connects the outer edge ($r_{\rm edge}$) and horizon ($r_{\rm h}$).

\subsection{Critical point analysis}

We analyze the dependency of the critical points on flow properties $i.e.,$ energy ($\mathcal{E}$) and angular momentum ($\lambda$) in the absence and presence of various dark matter distribution ($M_{\rm halo}=10 M_{\rm BH}, a_0=10^5 M_{\rm BH}$). In \cref{e_rc_lm}, the surface plot represents the nature of the critical points with $\mathcal{E}$ and $\lambda$. Critical points $r_c$ are plotted along the X-axis in the logarithmic scale, $\lambda$ is plotted along the Y-axis and $\mathcal{E}$ is plotted along
the Z-axis. The yellow, grey, and green colours denote the saddle, nodal, and spiral-type critical points. The range of critical points is not the same for all DM distributions and exhibits notable deviations from the Schwarzschild case for the HQ-type DM spike. A common feature is that there are upper and lower limits of energy and angular momentum, beyond which the multiple saddle-type critical points cease to exist. This happens due to the choice of the relativistic equation of state and is in agreement with the findings of previous studies~\cite{Dihingia:2020xxc}. As we increase the angular momentum, the likelihood of forming outer critical points is reduced. Under such conditions, the flow enters the BH only through the inner critical points.  Accretion flow with lower specific angular momentum passes through the outer critical points only. As these solutions are adiabatic in nature, the entropy accretion rate is constant throughout the journey from the outer edge to the horizon, passing through a critical point. 
\begin{figure*}[htbp!]
\includegraphics[width=0.32\textwidth]{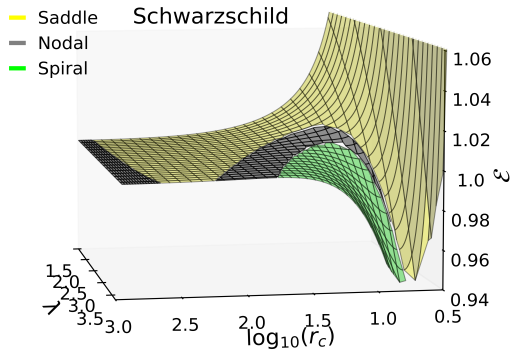}
\includegraphics[width=0.32\textwidth]{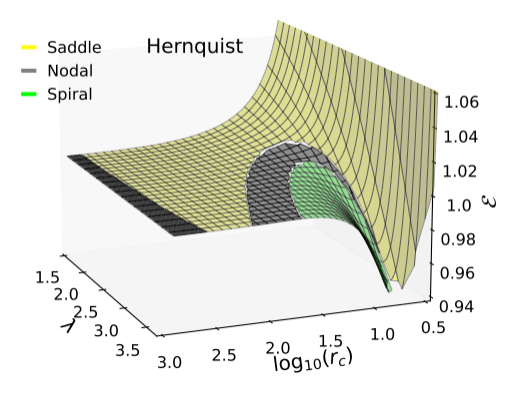}
\includegraphics[width=0.32\textwidth]{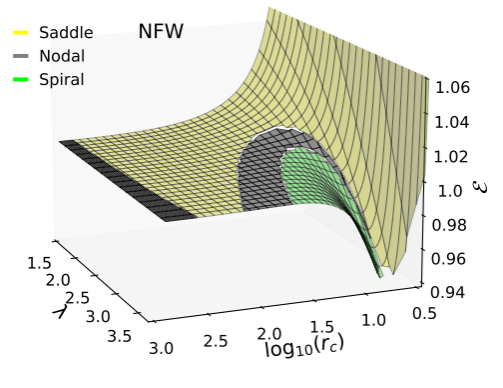}
\includegraphics[width=0.32\textwidth]{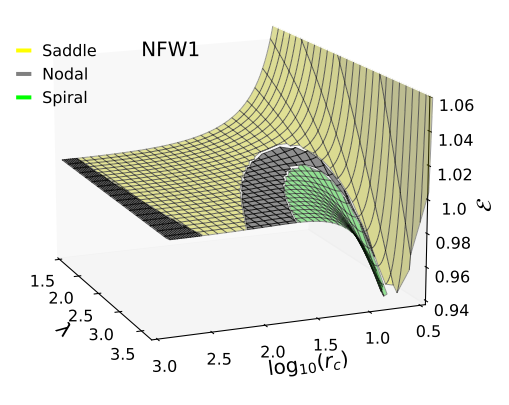}
\includegraphics[width=0.32\textwidth]{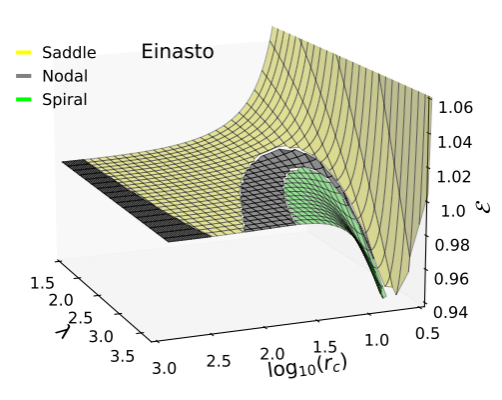}
\includegraphics[width=0.32\textwidth]{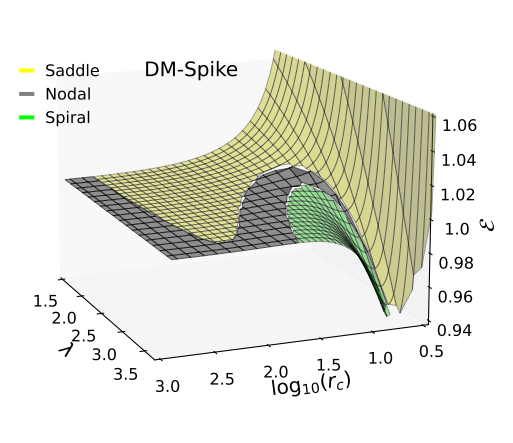}
	\caption{Variation of specific energy ($\mathcal{E}_c$) measured at a critical point $(r_{\rm c})$ with specific angular momentum ($\lambda_c$) for different dark matter density profiles with fixed $M_{\rm halo}=10 M_{\rm BH}$ and $a_0=10^5 M_{\rm BH}$. The top left plot corresponds to that of a  Schwarzschild BH. The yellow, grey, and green points represent the saddle, nodal, and spiral-type critical points. See the text for details.
    }
	\label{e_rc_lm}
\end{figure*}
To analyse the effect of $M_{\rm halo}$ and $a_0$, we refer the reader to \cref{sec:param}, where we focus on the portion of the $M_{\rm halo}-a_0$ parameter space for which both the inner and outer critical points exist.

\subsection{Global accretion solution}\label{sec:soln-topo}

In this section, we analyze the variation of flow topologies due to the presence of DM halo. Table~\ref{tab:topo} summarizes the variation of the flow topologies with specific angular momentum ($\lambda$) for $\mathcal{E} = 1.0015$, $M_{\rm halo}=10 M_{\rm BH}$, and $a_0=10^5 M_{\rm BH}$ for different types of DM distribution and equal mass Schwarzschild BH. For simplicity, in \cref{fig:Solution}, we depict the flow topologies around a vacuum Schwarzschild BH  and that around a galactic BH of the same mass immersed in an Einasto-type DM halo for the parameters mentioned above.
We observe that for the chosen specific flow energy, the specific angular momentum range for the I-type solution for the Schwarzschild BH falls within the  $\lambda$ range of the I-type solutions for all dark matter profiles. 
From \cref{fig:Solution}, we note that the presence of the Einasto-type DM halo changes the flow topology from O-type to A-type for $\lambda=2.875$ and A-type to W-type for $\lambda=3.1$. Increasing the angular momentum as $\lambda=3.36$, the topology of the solution changes from W-type to I-type in the presence of the Einasto-type DM distribution. Thus, changing the DM distribution can, in general, change the topology of the accretion flow depending on the choice of the DM and flow parameters. To elucidate this point further, in the following sections, we study the DM and flow parameter space for the existence of multiple saddle-type critical points.

\begin{figure*}[htbp!]
	\includegraphics[width=0.85\textwidth]{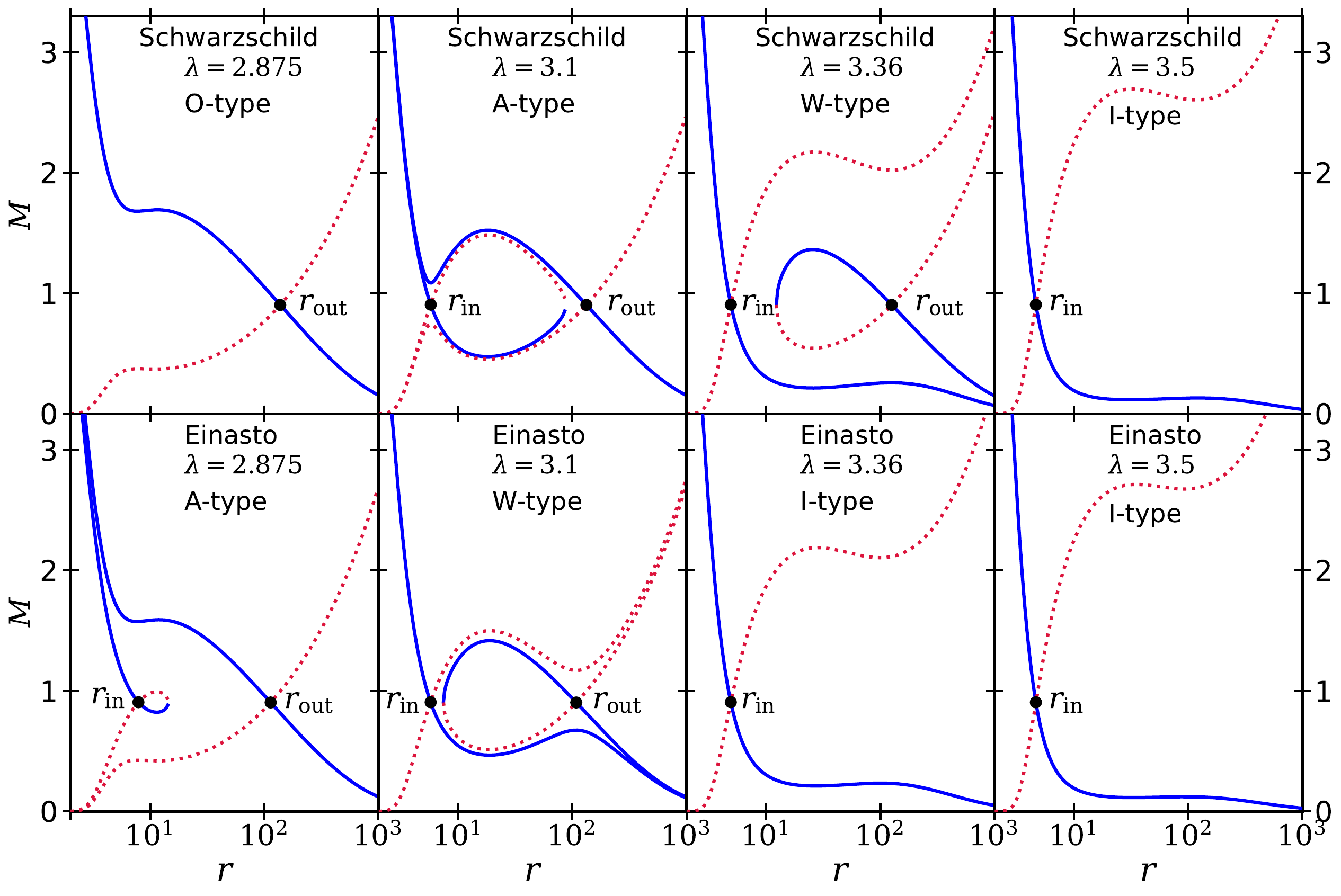}
	\caption{Accretion flow topologies around a galactic BH surrounded by an Einasto type DM distribution (bottom row), compared to an equal mass Schwarzschild BH (top row) with same flow parameters. Here, we choose $M_{\rm halo}=10 M_{\rm BH}$, $a_0=10^5 M_{\rm BH}$, $\mathcal{E}=1.0015$. See the text for details.
    }
	\label{fig:Solution}
\end{figure*}

\section{Parameter space for multiple critical points}\label{sec:param}
As discussed earlier, the occurrence of multiple (saddle-type) critical points happens only for a specific range of the flow parameters and DM parameters. In this section, fixing the DM parameters, we first scan the $\mathcal{E}-\lambda$ plane for multiple saddle type critical points~(Fig.\ref{fig:e_lm}), then fixing the flow parameters, we repeat the analysis in the $a_0-M_{\rm halo}$ plane (Fig. \ref{fig:Mh-a0}). Even within the portion of the parameter space allowing multiple (saddle-type) critical points, the topology of the accretion flow can be different. Though the entropy accretion rate ($\dot{\mathcal{M}}$) remains constant along the flow streamline, $\dot{\mathcal{M}}$ is different for the inner and outer critical points. Thus, we differentiate the accretion flow based on the solution topologies and the entropy accretion rates.
In Fig. \ref{fig:e_lm}, we show the $\mathcal{E}-\lambda$ parameter space for multiple saddle-type critical points for galactic BHs with different dark matter profiles. In each panel, we portray the corresponding parameter space for a Schwarzschild BH (black dotted line) of the same mass for comparison. The bounded regions with green (solid), blue (dash-dotted), and magenta (dashed) lines stand for three sets of DM parameters ($M_{\rm halo}, a_0$) = \{($10 M_{\rm BH}, 10^4 M_{\rm BH}$), ($10 M_{\rm BH}, 10^5 M_{\rm BH}$), ($10^2 M_{\rm BH}, 10^4 M_{\rm BH}$)\}, respectively. On the top and left of the boundary of the existence of multiple saddle critical points, the flow topologies are I-type and O-type, respectively. 
Contrary to the Hernquist, NFW and NFW1 profiles, the accretion flow does not possess multiple saddle type critical points for the HQ-type DM spike distribution with ($M_{\rm halo}, a_0$)=($10 M_{\rm BH}, 10^4 M_{\rm BH}$), whereas, for the Einasto type DM distribution the same is also true for ($M_{\rm halo}, a_0=10^2 M_{\rm BH}, 10^5 M_{\rm BH}$). In ~\cref{fig:e_lm}, we observe that compared to a vacuum Schwarzschild BH, the presence of DM halo shifts the allowed region of the parameter space for multiple (saddle) critical points to lower specific energies and smaller specific angular momenta. In cases of Hernquist, NFW, and NFW1 profiles, we observe that for a given halo-compactness ($\Psi=M_{\rm halo}/a_0$), increasing the halo mass shifts the upper bound of the specific energy slightly towards higher specific angular momentum. In all cases, we observe that for a fixed halo mass, increasing the halo compactness (lowering $a_0$) shrinks the allowed region of the parameter space, which also shifts towards lower specific angular momenta and lower specific energy of the flow. Although the admissible regions of the parameter space for the Hernquist and NFW distributions are almost similar, they are considerably different for the NFW1, Einasto and HQ-type DM spike profiles. The maximum values of $\lambda$ for all the cases considered in \cref{fig:e_lm} are set by the minimum value of the Keplerian angular momentum (i.e., $\lambda$ at the marginally stable orbit, See \cref{app:kep} for details) for the corresponding DM distributions.  In each panel, the shaded regions of the allowed parameter space correspond to the W-type solution, whereas the unshaded regions represent the A-type solution. Note that, for the NFW1 profile with $\Psi=0.001$, there exists a small portion of the $\mathcal{E}-\lambda$ parameter space, where only A-type solutions exist.
\begin{figure*}[htbp!]
	\includegraphics[width=0.49\textwidth]{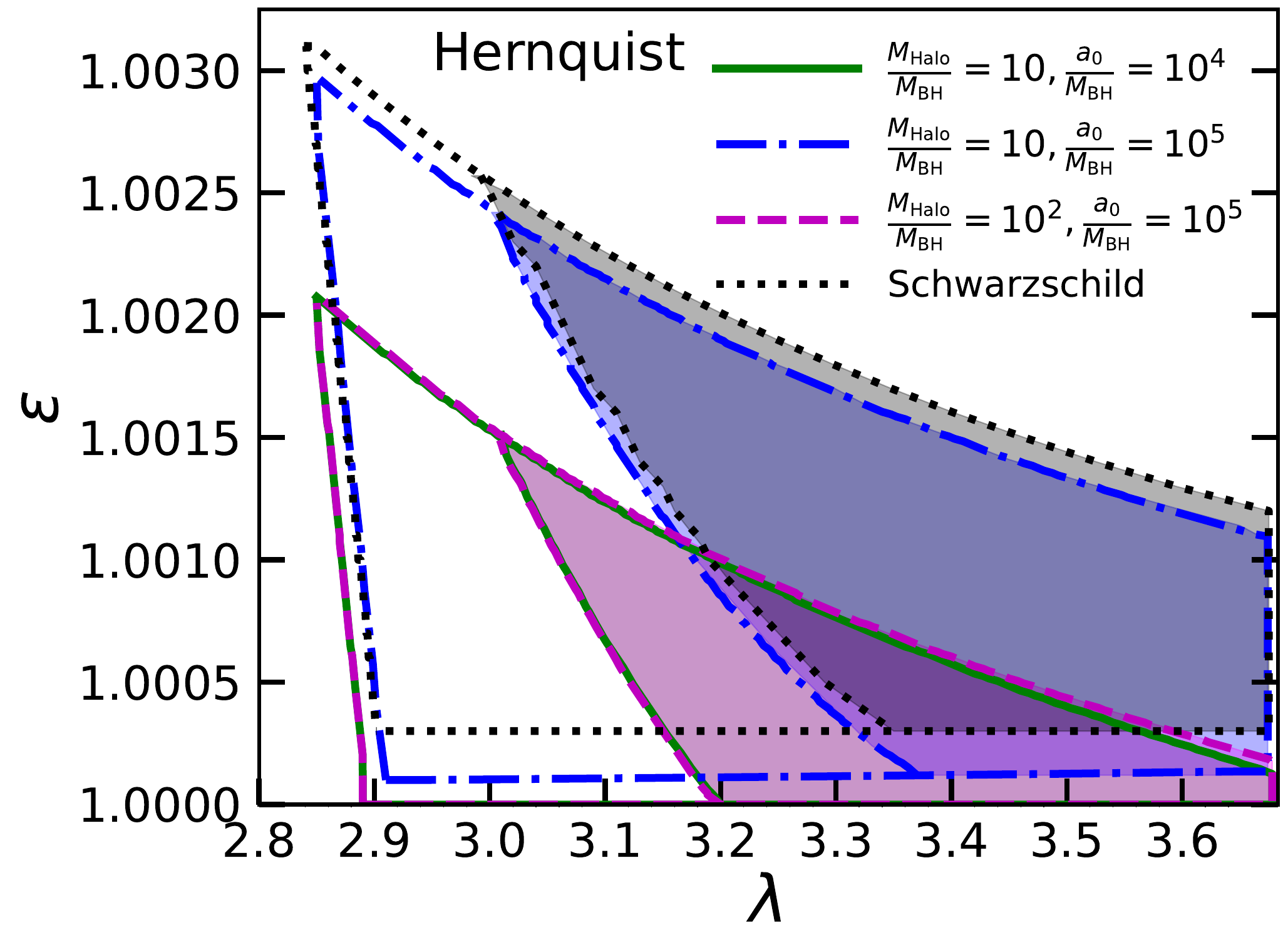}
	\includegraphics[width=0.49\textwidth]{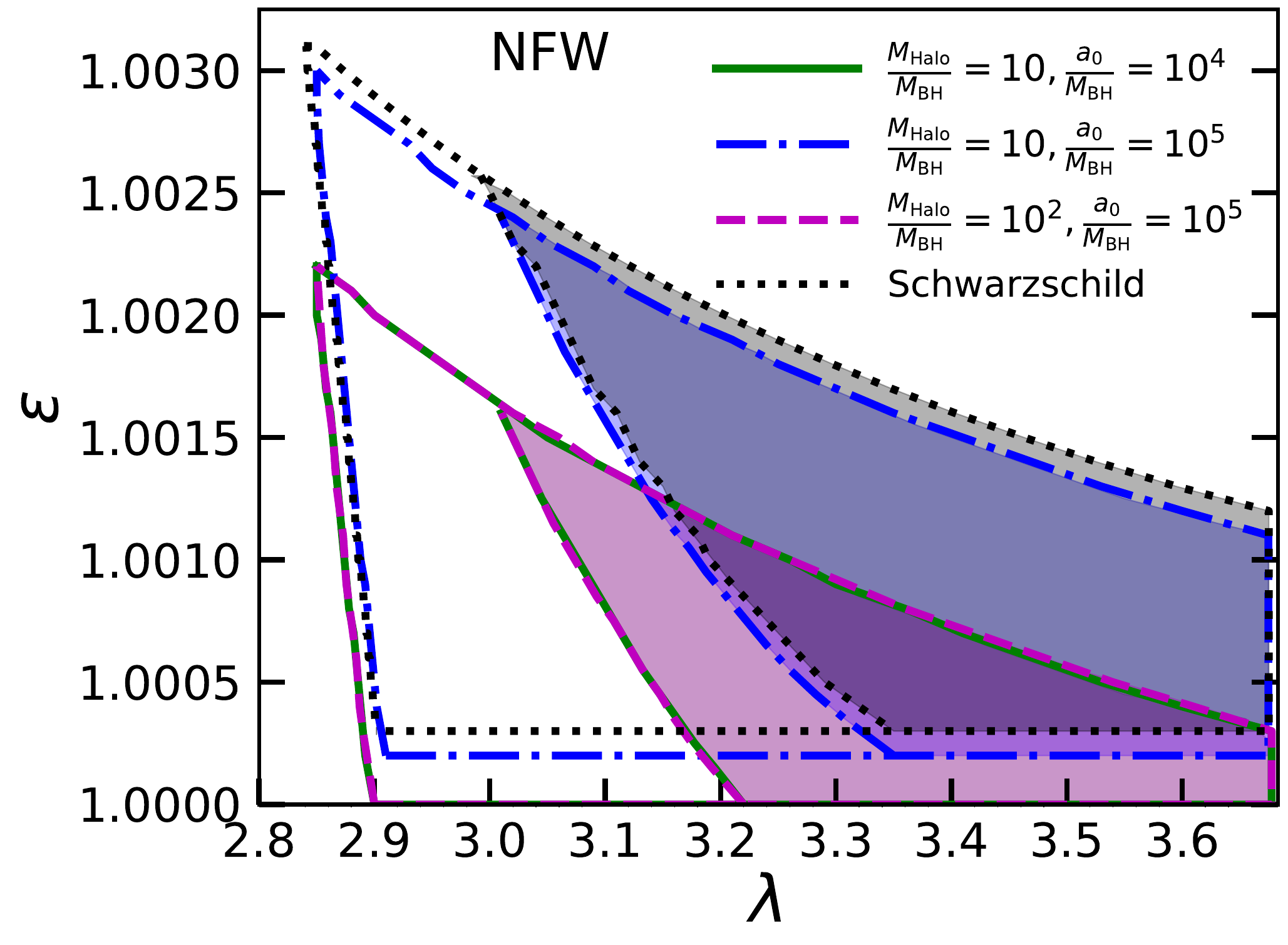}
	\includegraphics[width=0.49\textwidth]{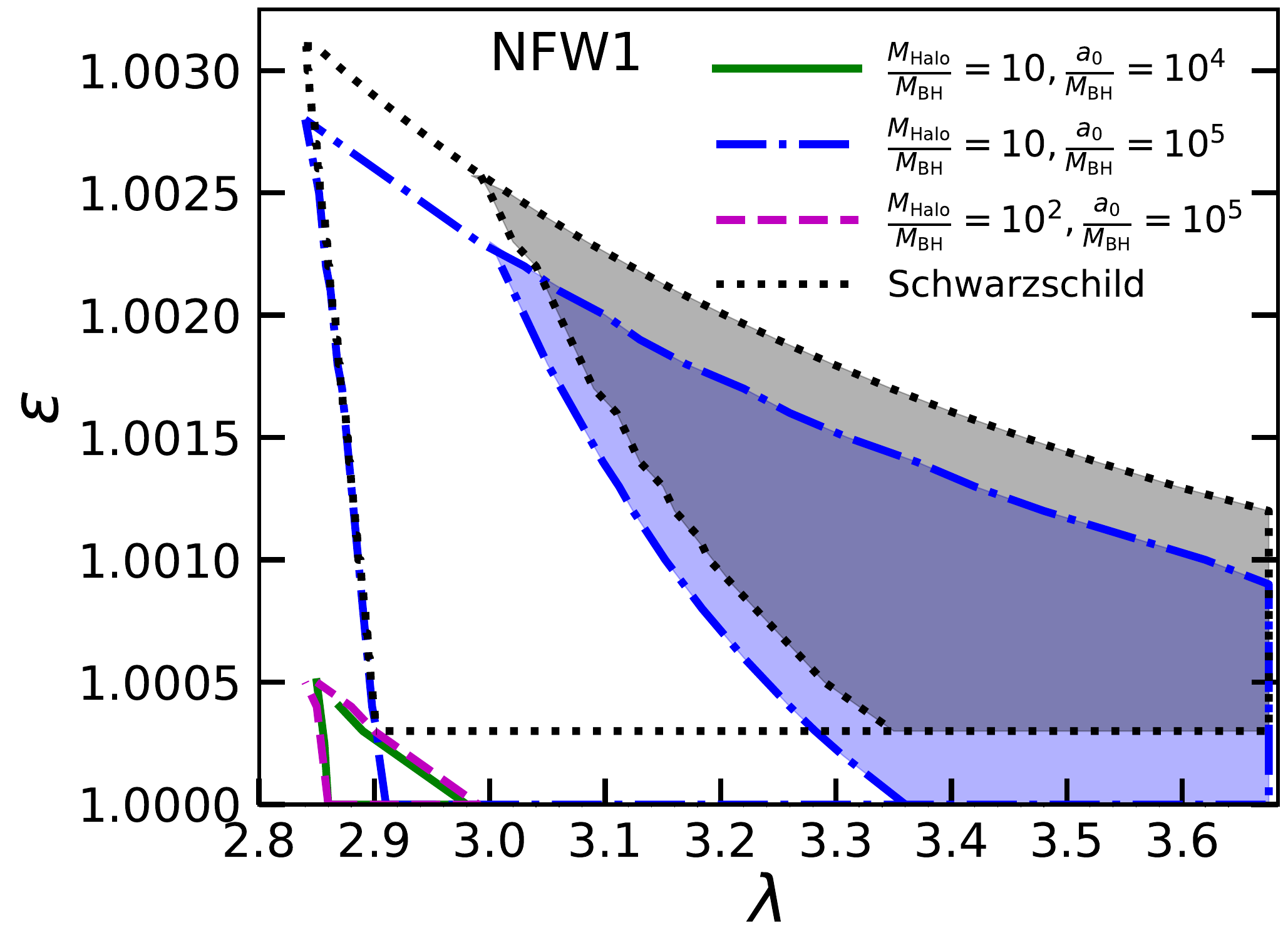}
    \includegraphics[width=0.49\textwidth]{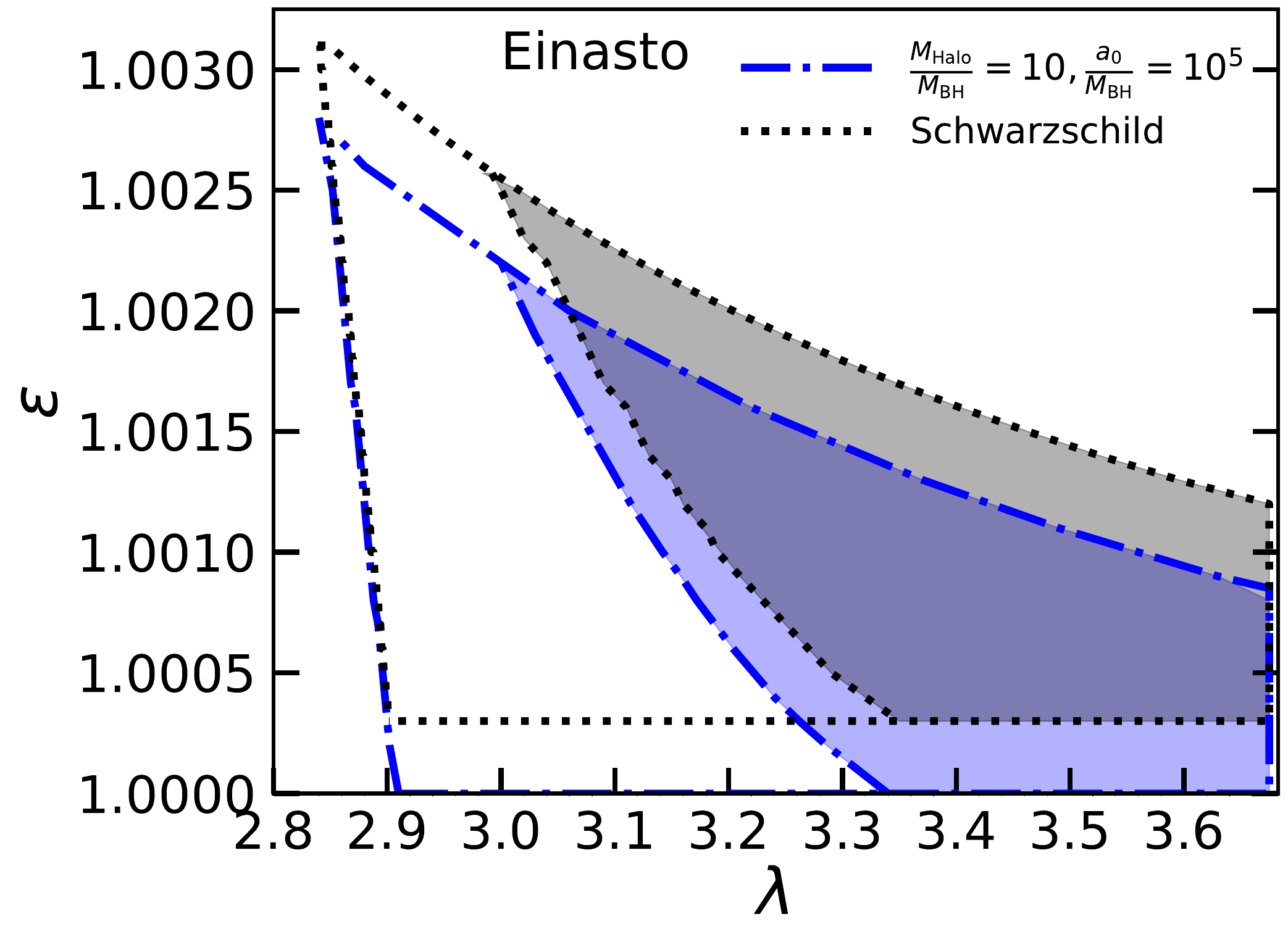}
    \includegraphics[width=0.49\textwidth]{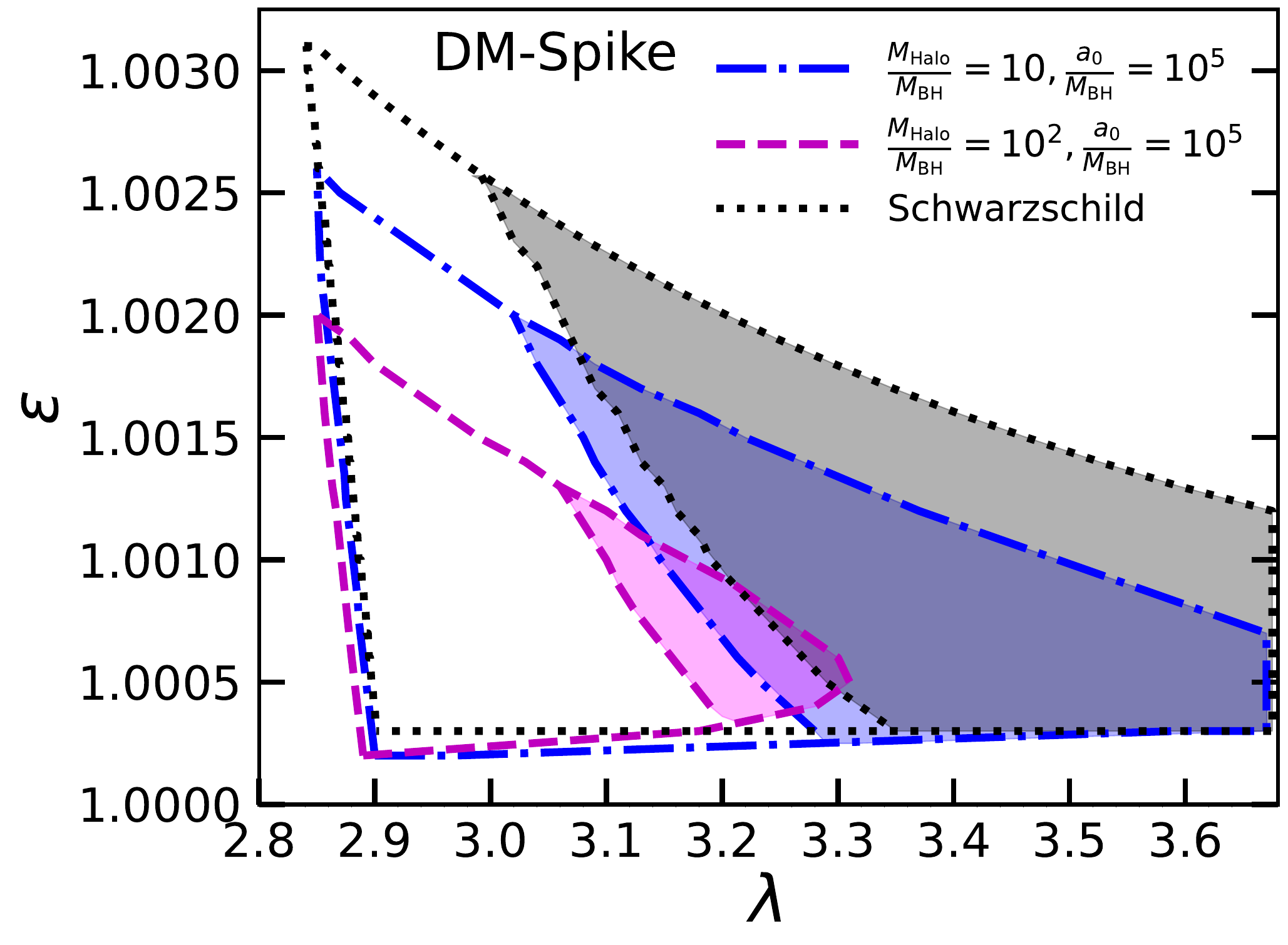}
	\caption{parameter space in $\mathcal{E}-\lambda$ plane for different dark matter profiles.  Here, we take three different ($M_{\rm halo}-a_0$) sets. The boundary region represents the parameter space in the $\mathcal{E}-\lambda$ plane for multiple saddle-type critical points. In all the figures, the black dotted line corresponds to a vacuum Schwarzschild black with the same mass. The middle lines ($\dot {\cal M}_{\rm in}=\dot {\cal M}_{\rm out}$) in each plot divide the A-type and W-type solutions. See the text for details.}
	\label{fig:e_lm}
\end{figure*} 

  \cref{fig:Mh-a0} illustrates the admissible region for multiple critical points in $a_0-M_{\rm halo}$ parameter space, corresponding to various sets of flow parameters, namely ($\mathcal{E},\lambda$)=\{(1.0009,3.0), (1.0009,3.5), (1.0022,3.0)\}. The magenta (solid), blue (dashed) and green (dash-dotted) lines represent the lower boundary for the admissible portion of the $a_0-M_{\rm halo}$ plane.  We observe that for a fixed set of flow parameters, the allowed region of the $a_0-M_{\rm halo}$ parameter space may contain either A type, W type or both types of accretion flow. For $\mathcal{E}=1.0022$ and $\lambda=3.0$, the magenta (dotted) line marks the upper bound of the W-type solutions (and lower bound of A-type solutions) for the NFW, NFW1 and Einasto DM distributions. The other solution topologies are marked in the figures respectively.

\begin{figure*}[htbp!]
\includegraphics[width=0.495\textwidth]{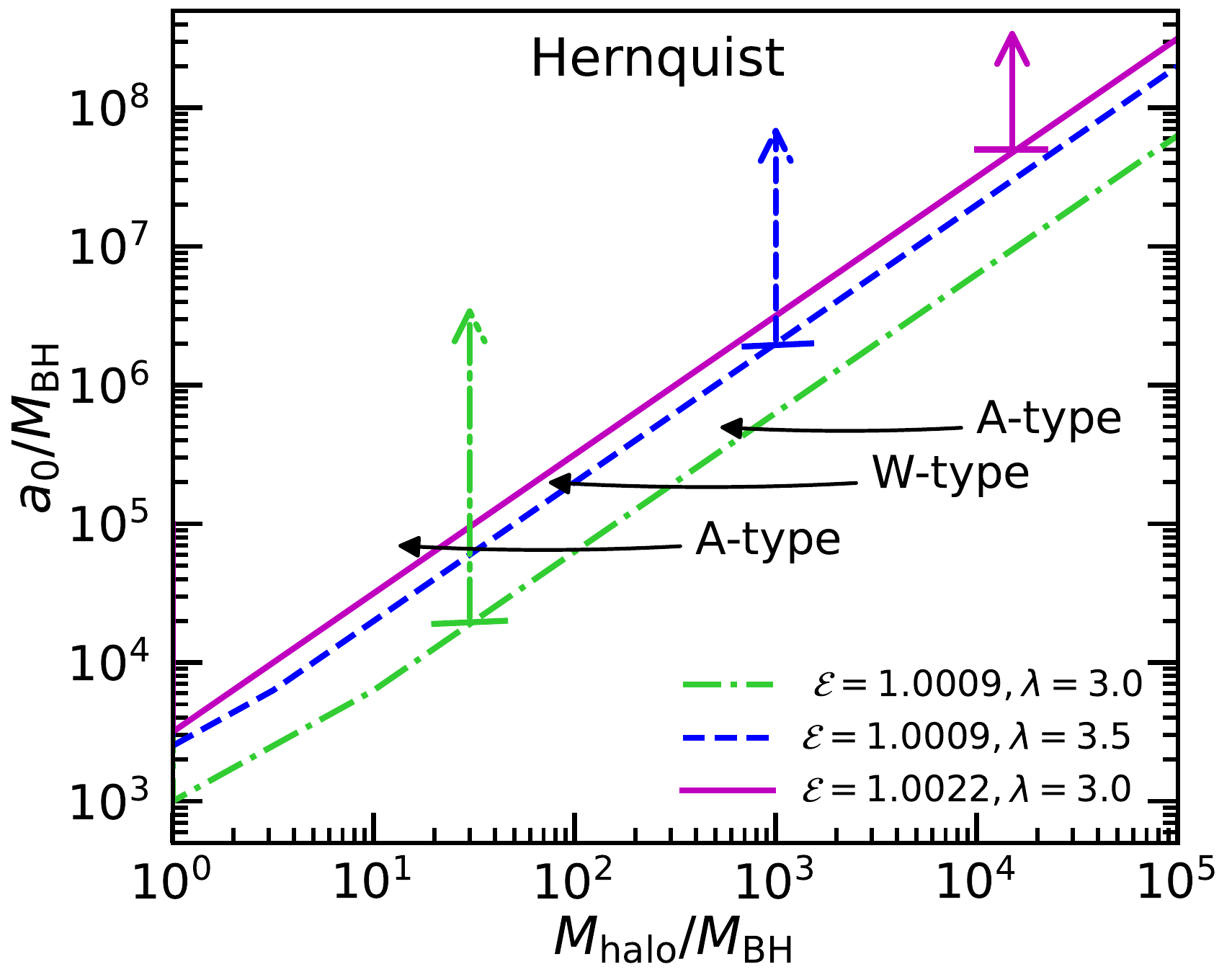}
\includegraphics[width=0.495\textwidth]{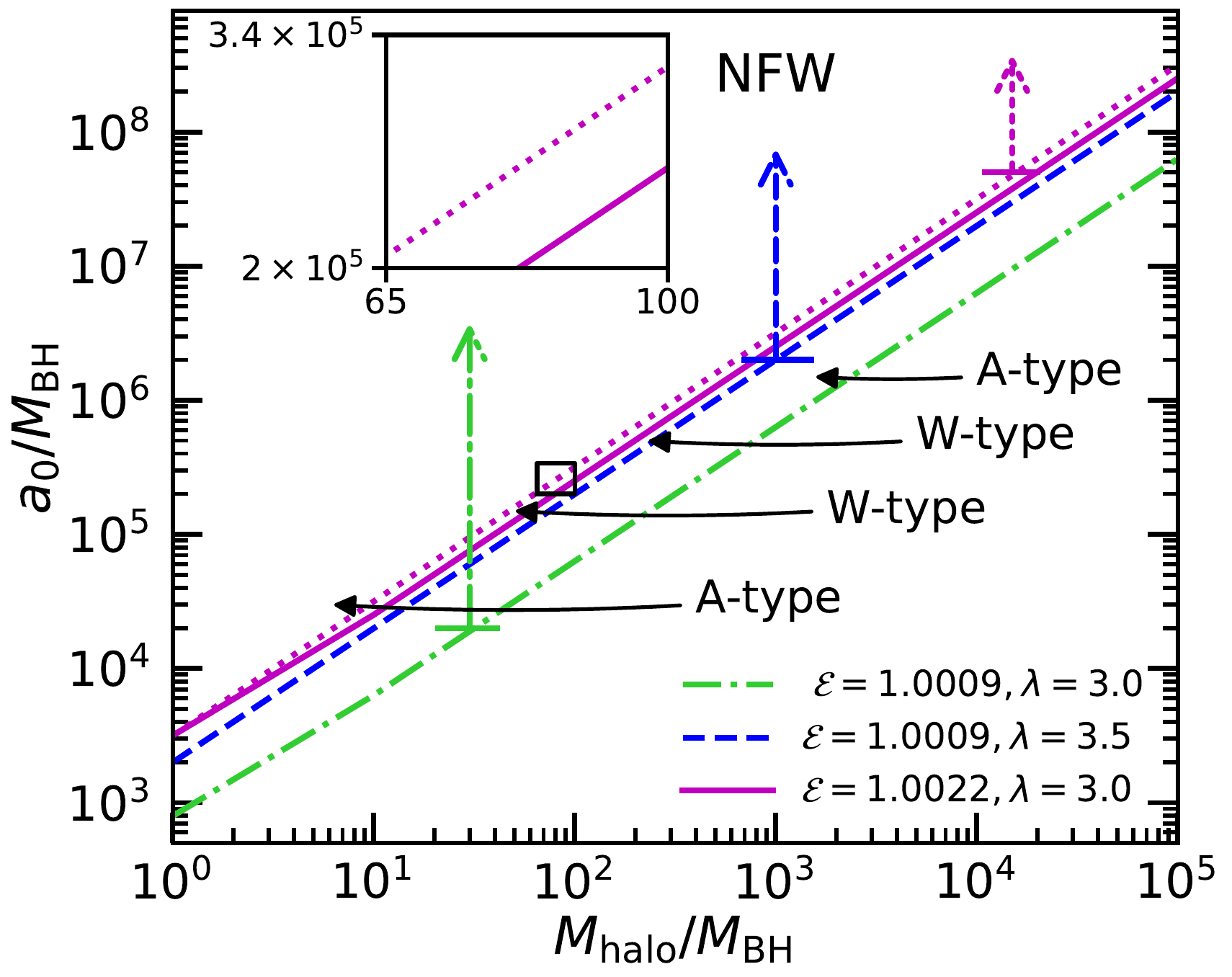}
\includegraphics[width=0.495\textwidth]{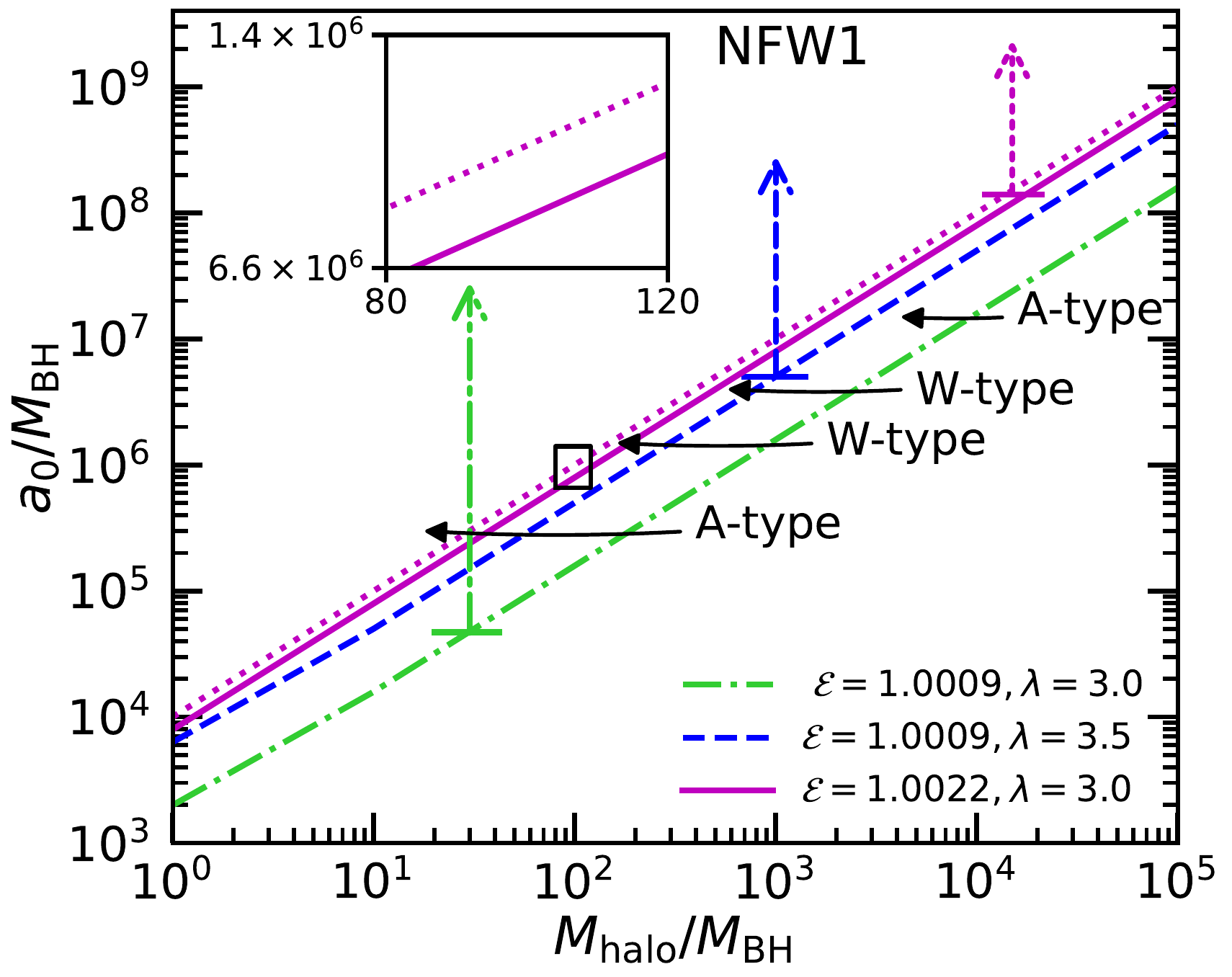}
\includegraphics[width=0.495\textwidth]{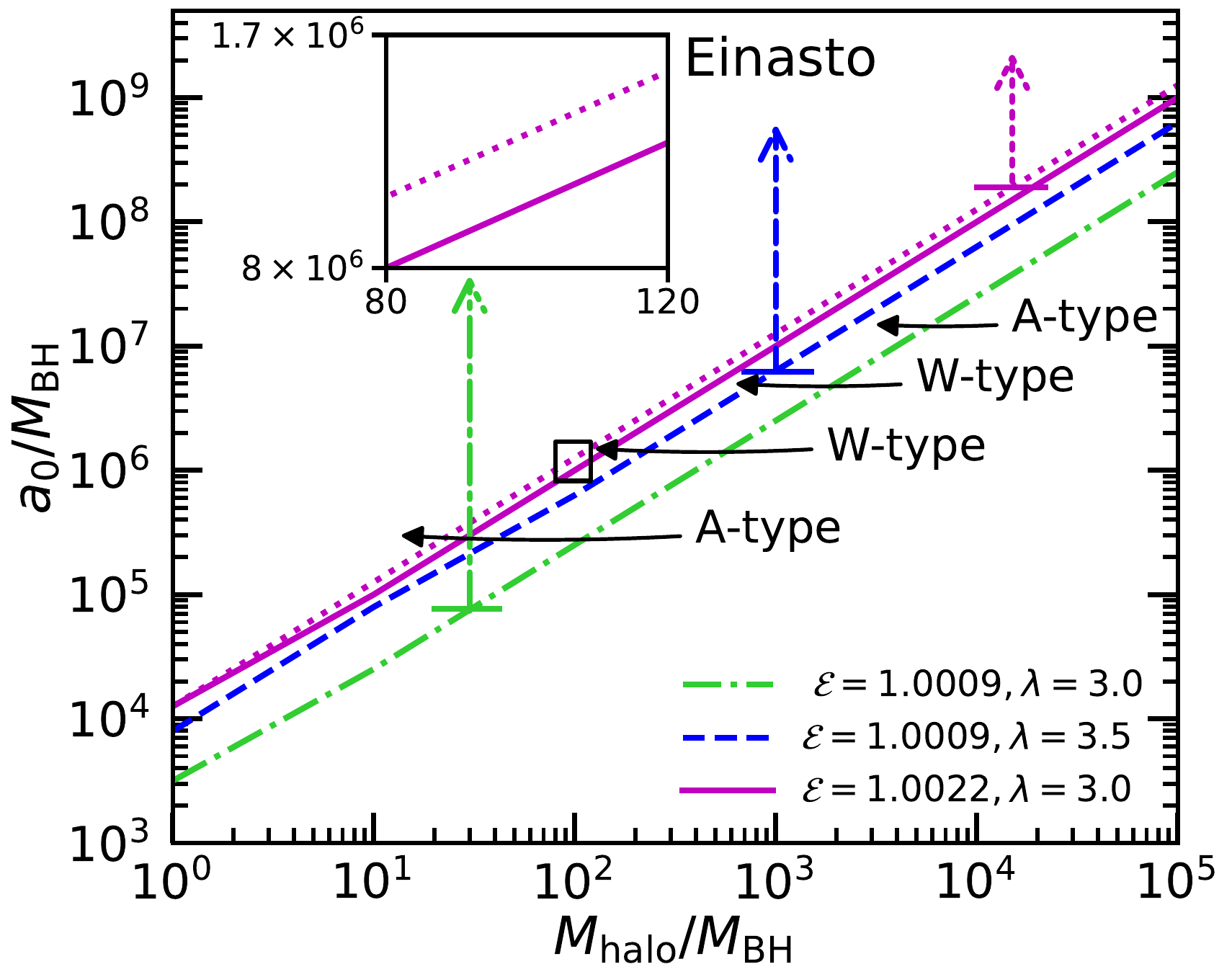}
\includegraphics[width=0.495\textwidth]{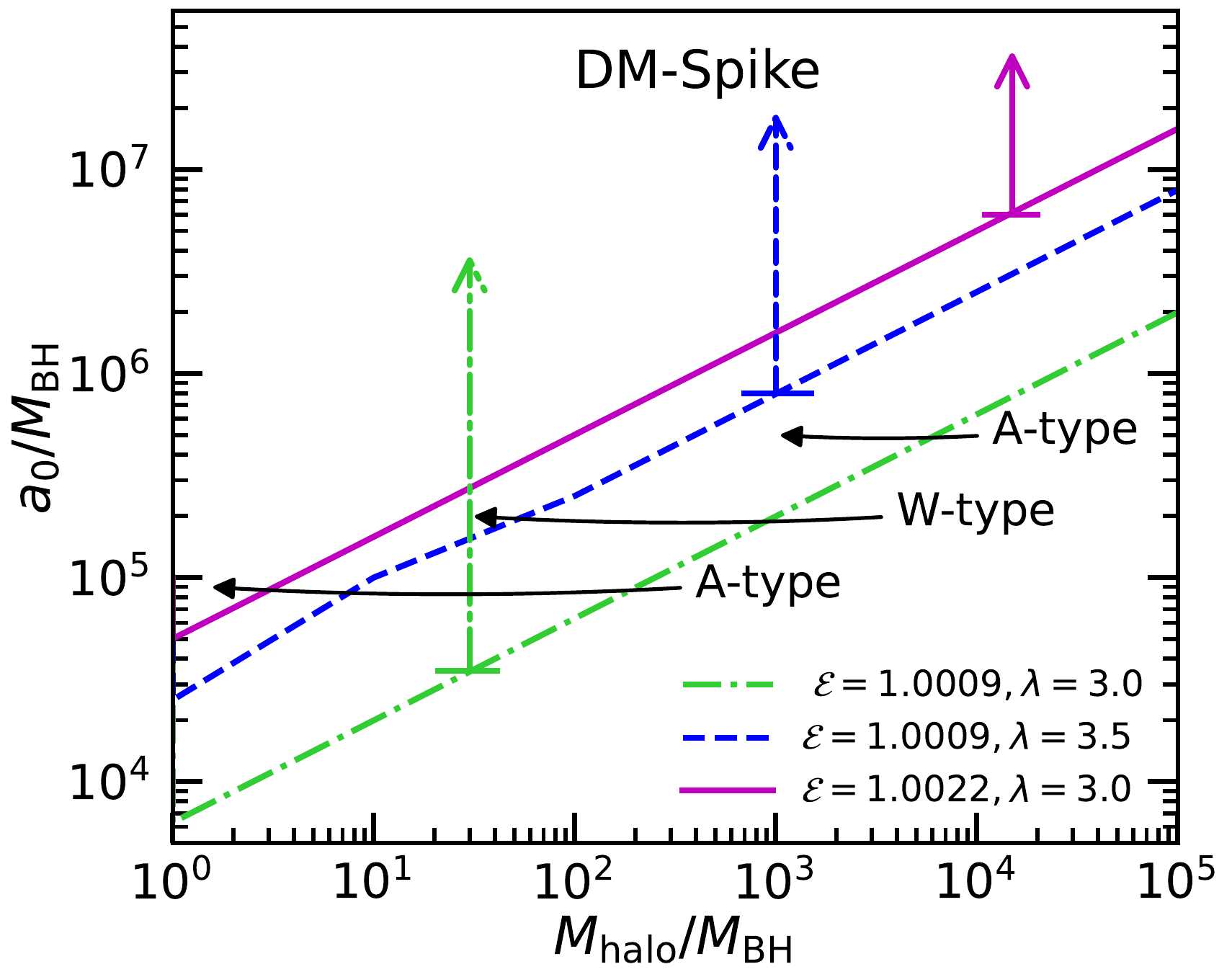}
    \caption{Parameter space in $a_0-M_{\rm halo}$ plane for different dark matter profiles.  Here we take three different ($\mathcal{E},\lambda$) combinations. In all the figures, we show the lower boundary of the allowed region of parameter space for multiple saddle-type critical points. See the text for details.}
    \label{fig:Mh-a0}
\end{figure*}
\begin{table*}[htbp!]
    \begin{tabular}{ccccc}
        \hline
           {}& $\lambda/\lambda_{K}$(O-type) & $\lambda/\lambda_{K}$(A-type) & $\lambda/\lambda_{K}$(W-type) & $\lambda/\lambda_{K}$(I-type) \\
           \hline
             Schwarzschild &<0.78  \quad &0.78--0.85\quad &0.85--0.94\quad & >0.94\\
             Hernquist &<0.78\quad &0.87--0.84 \quad &0.84--0.92 \quad &>0.92\\
             NFW &<0.78\quad &0.78--0.85 \quad &0.85--0.93\quad &>0.93\\
             NFW1 &<0.78\quad &0.78--0.84\quad &0.84--0.90\quad &>0.90\\
             Einasto &<0.78\quad &0.78--0.84 \quad &0.84--0.89\quad &>0.89\\
             DM - Spike &<0.78\quad &0.78--0.84\quad &0.84--0.87\quad &>0.87\\
        \hline
        \end{tabular}
        \caption{Table showing the specific angular momenta range for different accretion flow topologies around a galactic BH with different types of DM distributions with $M_{\rm halo}=10 M_{\rm BH}$, $a_0=10^5 M_{\rm BH}$ and $\mathcal{E}=1.0015$. The maximum value of $\lambda$ for each profile is fixed by the corresponding Keplerian angular momentum $\lambda_{K}$, where, $\lambda_{K}^{\rm Sch}=3.67424$, $\lambda_{K}^{\rm HQ}=3.67460$, $\lambda_{K}^{\rm NFW}=3.6756$, $\lambda_{K}^{\rm NFW1}=3.67519$, $\lambda_{K}^{\rm Ein}=3.67549$ and $\lambda_{K}^{\rm DM-Spk}=3.67625$~.}
        \label{tab:topo}
    \end{table*}
    
\section{Solution Properties}\label{sec:solprop}

In this section, we study how the presence of different DM halo distributions affects the properties of the accretion flow onto the central BH.
Following the I-type accretion  solution\footnote{For other types of solutions, the similar variations also exist.}, in Fig. \ref{fig:solprop}, we show a representative plot of the flow properties for different halo mass and scale radius for fixed $\mathcal{E}~(=1.001)$ and $\lambda~(=3.674)$. In the first two columns in \cref{fig:solprop}, the halo compactness ($\Psi$) increases from $10^{-4}$ to $10^{-2}$ for a fixed $M_{\rm halo}=10^2 M_{\rm BH}$), whereas in the third column, $M_{\rm halo}$ is increased to $10^{10} M_{\rm BH}$ keeping the halo compactness fixed at $\Psi=0.01$. The first to fifth rows in each column represents the Mach number ($M=v/c_s$), temperature ($T$), density ($\rho$), aspect ratio ($H/r$), and optical depth ($\tau_{\rm eff}$), respectively. For comparison in each panel, we also plot the corresponding flow property around a vacuum Schwarzschild BH.
We observe that the presence of DM halo shifts the critical point to smaller radii and decreases the flow velocity  (as well as Mach number) close to the BH, in general. As the mass accretion rate of the flow is assumed to remain constant throughout, the local mass density varies inversely with the flow velocity (see  \cref{eq:Mdot}). So, when the flow velocity (as well as Mach number) is lower than the Schwarzschild BH, the density becomes higher and vice versa. The constant specific energy of the accretion flow with reduced flow velocity results in increased thermal energy of the accreting fluid. This raises the local temperature resulting in an increased thermal pressure, which puffs up the disk, making it quasi-spherical. Thus, at a fixed radius, the flow with higher temperature has a higher aspect ratio~($H/r$).

Our analysis assumes the accretion disk to be optically thin. To ascertain that the accretion disk continues to remain optically thin, we explicitly calculate the optical depth ($\tau$) of the accretion disk. In fact, the optical thinness of the accretion disk is necessary for the emitted Bremsstrahlung radiation to be observable. Opacity to Bremsstrahlung radiation primarily occurs due to multiple coherent Thompson scattering of the emitted photon by the free electrons in the plasma as well as the free-free absorption in the fully ionized plasma. The opacity coefficient of Thompson Scattering is taken as $\kappa_{\rm s}=0.4 \rm{cm}^2 \rm{gm}^{-1}$. Following~\cite{maoz}, we evaluate the Rosseland mean opacity coefficient for free-free emission as,
\begin{equation}\label{eq:kff}
\begin{split}
k^{\rm ff}_{\rm R} = &0.64 \times10^{23} \rho T_e^{-7/2} \\
&\times(1+4.4\times10^{-10}T_e)\bar{g}_{\rm ff}~\rm{cm}^2 \rm{gm}^{-1}.
\end{split}
\end{equation}
Considering the half-thickness ($H$) of the disk as the typical associated length scale, we evaluate the effective optical depth as,
\begin{equation}\label{eq:tau}
    \tau_{\rm eff}=\sqrt{\tau_a\left( \tau_a +\tau_s \right)}~,
\end{equation}
where $\tau_a=k_{\rm s} \rho H$ is the optical depth due to Thompson scattering and $\tau_a=k^{\rm ff}_{\rm R} \rho H$ is the absorption optical depth. We note from  \cref{eq:kff,eq:tau}  that the optical depth depends on the local flow temperature, density and disk height. Since the presence of DM halo changes each of these flow properties, the exact deviation of the local optical depth is the result of an interplay between them.  However, we note from \cref{fig:solprop} that the presence of DM halo does not make the accretion disk optically thick. The disk is particularly optically thinner close to the inner edge from where maximum emission occurs (see \cref{sec:lumi} for details).

Comparing the first and second columns, we note that increasing the halo compactness increases the deviation from the Schwarzschild BH with the Einasto profile showing the maximum variation.
Apart from the HQ-type DM spike profile, increasing the halo mass also increases the deviation of the flow properties from those of the Schwarzschild BH. In the case of HQ-type DM spike distribution, the density profile has a strong nonlinear dependence on the halo mass.  For this particular case, it is observed that for small $M_{\rm halo}$ the DM density at a fixed radius decreases with the halo mass, however for large $M_{\rm halo}$, the behaviour is reversed. This change in the halo density with $M_{\rm halo}$ strongly influences the background geometry, which in turn affects the effective potential, causing the flow properties to behave nonlinearly with the halo mass (see \cref{app:DMspk} for details on HQ-type DM spike profile).
 
\begin{figure*}[htbp!]
\includegraphics[width=1\textwidth]{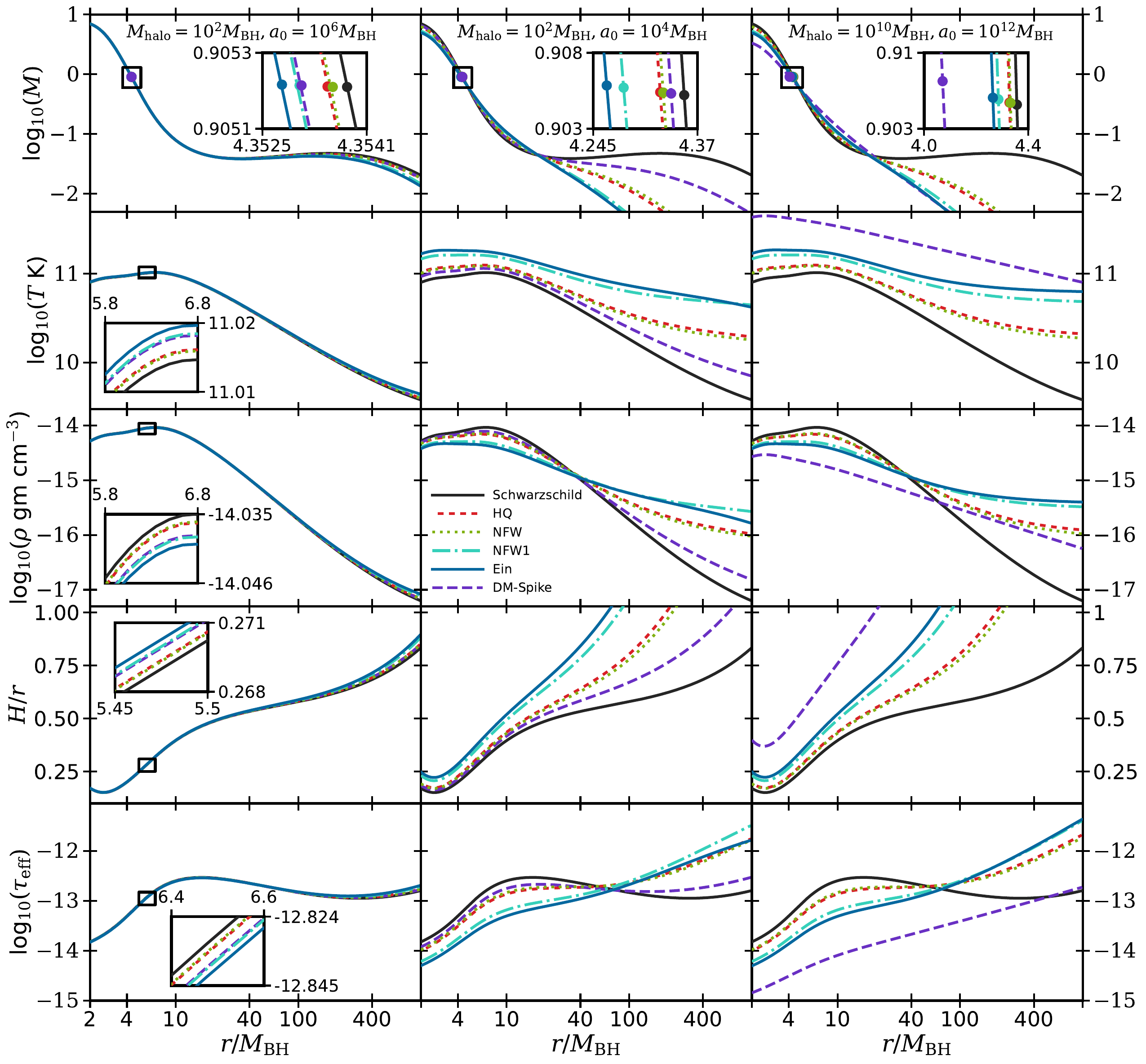}
    \caption{Variation of flow variables of I-type accretion flow with radial coordinate ($r$) for different compactness and halo mass with $\mathcal{E}=1.001$ and $\lambda=3.674$. In the first column, we consider $M_{\rm halo}=10^2M_{\rm BH}$ and $a_0=10^6M_{\rm BH}$. In the second and third columns, plots are for ($M_{\rm halo}$, $a_0$)= ($10^2M_{\rm BH},10^4M_{\rm BH}$) and ($10^{10}M_{\rm BH},10^{12}M_{\rm BH}$), respectively. See the text for details.} 
     \label{fig:solprop}
\end{figure*}

\section{Radiative emission properties}\label{sec:lumi}

In this section, we study the emission of the thermal Bremsstrahlung (free-free) radiation from the accretion disk. Bremsstrahlung emission is the radiation emitted from an accelerating electron as it passes through the field of a heavy ion. However, in the case of hot accretion flow (HAF), the flow is thermally relativistic ($k_{B}  T_e>m_e c^2$), making the electron-electron emission more significant. Thus, following  \cite{Novikov-Thorne1973}, we incorporate the relativistic effect and electron-electron emission in addition to the standard electron-ion emission channel to obtain an approximate expression for Bremsstrahlung emissivity at a frequency $\nu$ as,
\begin{equation}\label{eq:emissivity}
\begin{split}
    \varepsilon_\nu=&\frac{32 \pi e^6}{3 m_e c^3}\sqrt{\frac{2 \pi}{3 m_e k_{B}}} n_e n_i z^2 T_e^{-1/2} \\
    &\times(1+4.4\times 10^{-10} T_e)e^{-h\nu/k_B T_e} \bar{g_b} ,
\end{split}
\end{equation}
where $h$ is the Planck constant, $k_B$ is the Boltzmann constant, $n_i$ is the ion-number density,  and $e$, $m_e$, $n_e$, and $T_e$ are the charge, mass, number density and temperature of electrons, respectively. The factor $\bar{g_b}$ is the thermally averaged Gaunt factor that incorporates the quantum-mechanical corrections and is taken to be 1.2~\cite{Yarza:2020loo}. Here, we consider the plasma to be hydrogenic and the electron number density to be equal to the ion number density ($n_e=n_i$). Furthermore, since the electrons are much lighter than the ions, the electron temperature must be lower than the ion temperature, which, in turn, is approximately equal to the flow temperature $T$. Thus, we approximate the electron temperature as $T_e\approx T/10$ throughout the accretion disk for simplicity~\cite{Yarza:2020loo}. The strong gravitation potential of the central BH implies that the emitted radiation is red-shifted as it travels towards an asymptotic observer. The rotation of the accretion disk also induces a Doppler shift of the emitted radiation. Thus, ignoring any light-bending effects, the frequency of the emitted radiation is related to the observed frequency at asymptotic infinity as~\cite{Luminet:1979nyg},
    \begin{equation}\label{eq:redshift}
        \nu_e=\nu_o (1+z)=\nu_o u^t \left(1+\frac{e \lambda f(r)}{r^2 c} \sin \theta_0 \sin\phi\right),
    \end{equation}
where $z$ is the redshift factor, and $\theta_0$ is the inclination angle of the accretion disk, which we consider to be $\pi /4$ for galactic BHs. Taking all these into account, we write the monochromatic disk luminosity for an asymptotic observer as,
\begin{widetext}
\begin{align}\label{eq:lumin}
L_{\nu_0}=&2\int_{r_{\rm h}}^{r_{edge}}\int_{0}^{2\pi} \varepsilon_{\nu_e} H r~dr d\phi ,\nonumber\\
=&4.875 \times10^{10} \bar{g_b}~{\rm{erg}~ \rm{s}^{-1} \rm{Hz}^{-1} } 
\int_{r_{\rm h}}^{r_{\rm edge}}\int_{0}^{2\pi} \left[ \rho^2 (T/10)^{-1/2}\times (1+4.4 \times 10^{-11}T) e^{-10(1+z) h \nu_o/k_B T} H r ~dr d\phi\right],
\end{align}
\end{widetext}
where $ r_{\rm h}$ and $r_{\rm edge}$ are the inner and outer edges of the accretion disk, respectively. For our analysis, we take $r_{\rm edge}$ = $\min(1000 M_{\rm BH},r_0)$, where $r_0$ is such that $H/r_0\sim 1$. This choice of $r_{\rm edge}$ is essential to ensure that our analysis respects the assumptions in section \ref{sec:assume}. 
To analyse the radiative properties of the galactic BHs, we assume the mass of the central supermassive BH as $M_{\rm BH}=5\times10^{8} M_\odot$~\cite{Boshkayev:2020kle, Wu:2004ry} and mass accretion rate of $\dot{M}=0.01 \dot{M}_{\rm EDD}$, where $\dot{M}_{\rm EDD}=1.39\times10^{18}\left(M/M_\odot\right)~\rm{gm~ s^{-1}}$ is the Eddington mass accretion rate.

\begin{figure*}[htbp!]
\includegraphics[width=\textwidth]{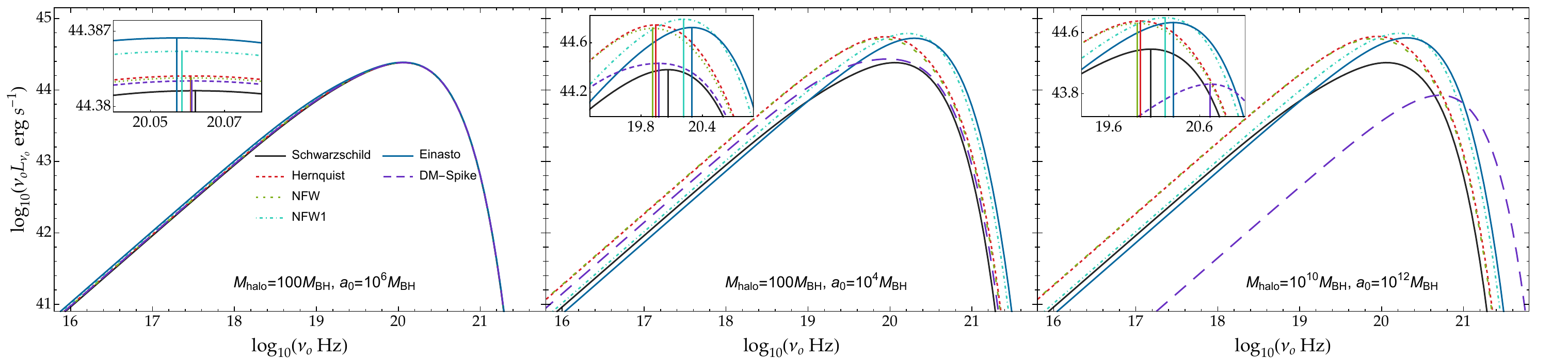}
\caption{Plot of the spectral energy distribution for different DM distributions for different values of the halo mass and scale radius: $ M_{\rm halo}=100 M_{\rm BH}, a_0=10^6 M_{\rm BH}$ (left), $ M_{\rm halo}=100 M_{\rm BH}, a_0=10^4 M_{\rm BH}$ (middle), $ M_{\rm halo}=10^{10} M_{\rm BH}, a_0=10^{12} M_{\rm BH}$ (right).  Halo compactness $(M_{\rm halo}/a_0)$ changes from $10^{-4}$ (left panel) to $10^{-2}$ (middle panel) for a fixed halo mass, whereas it is kept fixed at $0.01$ for right panel ($M_{\rm halo}=10^{10} M_{\rm BH}$) with $M_{\rm BH}=5\times10^8 M_\odot$. The solid black curve in each plot shows the SED for a Schwarzschild BH of the same mass. The insets show the location of the peak of the SEDs in each case. The flow parameters are $\lambda=3.674$ and $\mathcal{E}=1.001$. See the text for details.}
\label{fig:sed}
\end{figure*}

In \cref{fig:sed}, we show the spectral energy distribution (SED)  of galactic BHs for the different DM distributions with varying halo mass and compactness, ($M_{\rm halo}/M_{\rm BH}$, $\Psi$)=\{(100, $10^{-4}$), (100, 0.01), ($10^{10}$, 0.01) \}. We choose the specific energy and angular momentum of the flow as $\mathcal E=1.001$ and $\lambda=3.674$. We choose $\lambda < \lambda_{\rm K}^{\rm Sch}$ to ensure the flow remains sub-keplerian all throughout, where $\lambda_{\rm K}^{\rm Sch}$ is the Keplerian angular momentum for Schwarzschild BH (see \cref{app:kep} for details). We observe that the presence of DM causes significant deviation of the SEDs from that of a vacuum Schwarzschild BH of the same mass. The SED also shows a sharp cutoff at $\nu\approx 10^{21} -10^{22}$ Hz, which is governed by the electron temperature at the inner edge of the accretion disk. The presence of a DM halo increases the cutoff frequency from that of the vacuum Schwarzschild BH. Although qualitatively similar, the SEDs for a given DM distribution show significant quantitative variation with the halo mass and compactness.
For the low compactness of the DM halo, the spectral energy distribution at high frequency is almost similar to that of the vacuum Schwarzschild BH, whereas the deviation from the Schwarzschild BH is relatively large in the low frequencies. On the other hand, for very high compactness, SED shows an even stronger deviation from the vacuum Schwarzshild BH. For a given halo mass, as the halo gets more and more compact, the spectral luminosity increases both in the high and low-frequency regimes. The peak luminosity is also higher for more compact halos and occurs at a relatively lower frequency (except for the Einasto and NFW1 distributions, where it shifts to a higher frequency). The location of the peak of SED corresponds to the frequency and, hence, the energy of the electrons that provide the maximum contribution to the emitted radiation. The cutoff frequency is also higher for more compact halos, which is again an artefact of the higher flow temperature for more compact halos.
Though the changes in the SED with halo mass for a given halo compactness are even smaller, we observe the following general trend. The peak luminosity of the Hernquist, NFW, NFW1 and Einasto profiles slightly increases with the halo mass (left to right in \cref{fig:sed}) and shifts towards a higher frequency. However, for the HQ-type DM spike profile, the peak spectral luminosity decreases and shifts towards higher frequency as the halo mass is increased. For Hernquist and NFW profiles, the spectral luminosity increases with the halo mass in both low and high-frequency ranges. For NFW1, Einasto and HQ-type DM spike distributions, though the spectral luminosity at high frequencies increases with the halo mass, the behaviour is reversed at low frequencies. The cutoff frequency stays unaffected by the change in halo mass for a given halo compactness except for the HQ-type DM spike distribution, where it increases with the halo mass. 
Integrating the spectral luminosity over all frequencies, we get the bolometric disk luminosity as,
\begin{equation}
    \label{eq:bolo-lumin}
    L=\int_0^\infty L_{\nu_o} d\nu_o.
\end{equation}
\begin{figure*}[htbp!]
\includegraphics[width=0.9\textwidth]{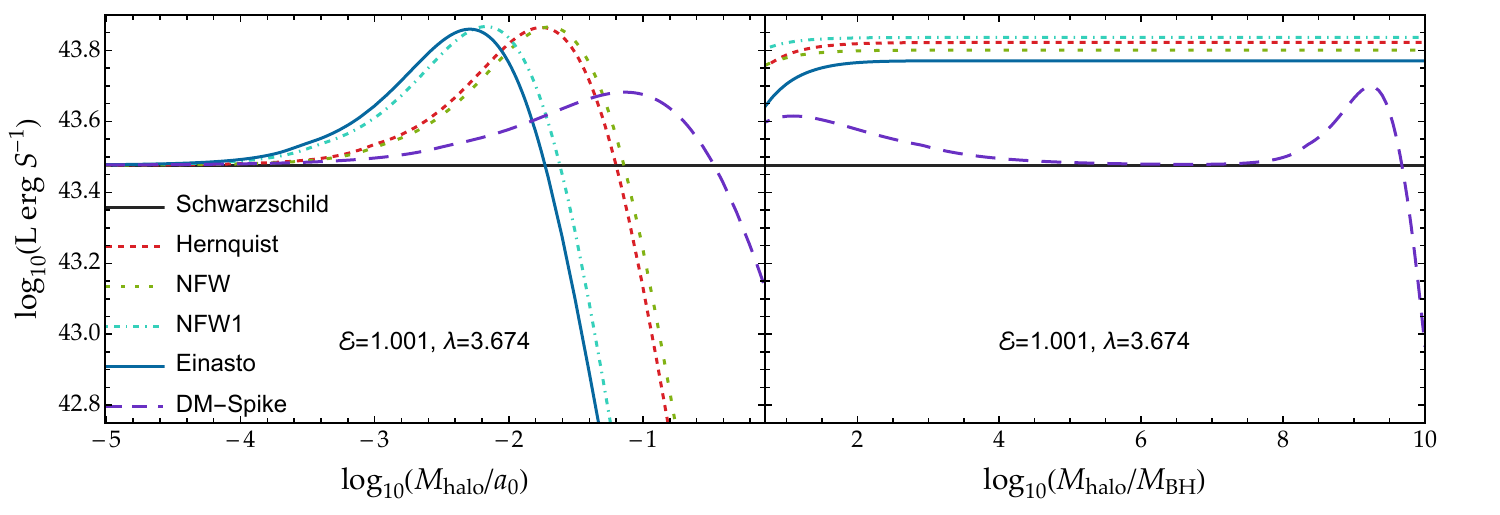}
    \caption{Right: Plot of bolometric luminosity for different dark matter profiles with halo compactness, $M_{\rm halo}/a_0$ for $M_{\rm halo}=100 M_{\rm BH}$ for $\mathcal{E}=1.001$ and $\lambda=3.674$. Left: Plot of the luminosity for different dark matter profiles with $M_{\rm halo}$ for $M_{\rm halo}/ a_0=0.01$ for the same values of energy and specific angular momentum of the flow. See the text for details.}
    \label{fig:plumi}
\end{figure*}

Figure~\ref{fig:plumi} shows the variation of the bolometric luminosity with the halo compactness and halo mass for a given value of the flow energy\footnote{As the $\mathcal{E}$ is increased keeping a constant $\lambda$ (=3.674), the difference with the Schwarzschild bolometric luminosity starts decreasing. At the value of the specific energy at which the Schwarzschild bolometric luminosity is maximum, all the DM distributions show a reduced bolometric luminosity.} and angular momentum. For constant halo mass, as the halo compactness increases, the luminosity for each DM distribution increases from the Schwarzschild value reaching a maximum and then decreasing sharply. The compactness at which the maximum bolometric luminosity is obtained is different for different DM distributions. The maximum value of the luminosity for the Hernquist, NFW, NFW1 and Einasto profiles is almost the same; however, for the DM spike profile, the maximum attained luminosity is substantially smaller. In \cref{fig:plumi}, we observe that for constant halo mass, as the halo compactness increases, the maximum bolometric luminosity for the Einasto distribution is obtained, followed by the NFW1, Hernquist, NFW and the HQ-type DM spike distributions. As discussed in \cref{sec:solprop},  when the compactness of the DM halo increases, the critical point shifts to smaller radii and the temperature of the accretion flow also increases, which in turn increases the luminosity. However, the increasing temperature also puffs up the accretion disk. Since our analysis is valid when the aspect ratio is less than unity ($H/r \leq1$), the effective portion of the accretion disk contributing to the luminosity reduces, decreasing the luminosity at high compactness. 

On the contrary, for fixed compactness, the bolometric luminosity increases with the halo mass, but it soon saturates with the NFW1 profile, giving the maximum luminosity followed by the Hernquist, NFW, Einasto and HQ-type DM spike distributions. The variation of the bolometric luminosity for the HQ-type DM spike distribution with the halo mass shows a rather curious behaviour when compared to other profiles (see the right panel of \cref{fig:plumi}). 
For sufficiently high halo mass, the bolometric luminosity of the HQ-type DM spike distribution is less than that of a vacuum Schwarzschild BH of the same mass.
This initial variation of the bolometric luminosity with the halo mass for a HQ-type DM spike distribution of constant compactness $(\sim 0.01)$ is related to the variation of the halo density profile and the associated change in geometry and properties of the accretion flow with the halo mass as discussed in \cref{sec:solprop,app:DMspk}. At a high enough halo mass, the relative disk height attains unity at a much smaller radius, reducing the luminosity even below the Schwarzschild limit.

 In \cref{fig:maxlumi}, we plot the maximum luminosity of the galactic BH with different DM profiles for a constant halo mass of $100 M_{\rm BH}$ and halo compactness~$0.01$  with $M_{\rm BH}=5\times 10^8 ~M_\odot$. In the plot, we vary the specific flow energy while maintaining the specific angular momentum of the flow slightly less than the Keplerian angular momentum in each case. We observe that the maximum bolometric luminosity is obtained for the NFW1 distribution, followed by the Einasto, Hernquist and NFW distributions. The maximum luminosity in each of these cases is higher than the maximum bolometric luminosity of a vacuum Schwarzschild BH. However, the bolometric luminosity of the HQ-type DM spike distribution around an equal mass BH with the same set of halo parameters shows a significantly lower luminosity. This low luminosity is a manifestation of the fact that at the chosen compactness and halo mass, the critical radius $r_0$ is very small (see \cref{fig:solprop}).
\begin{figure*}[htbp!]
\includegraphics[width=0.7\textwidth]{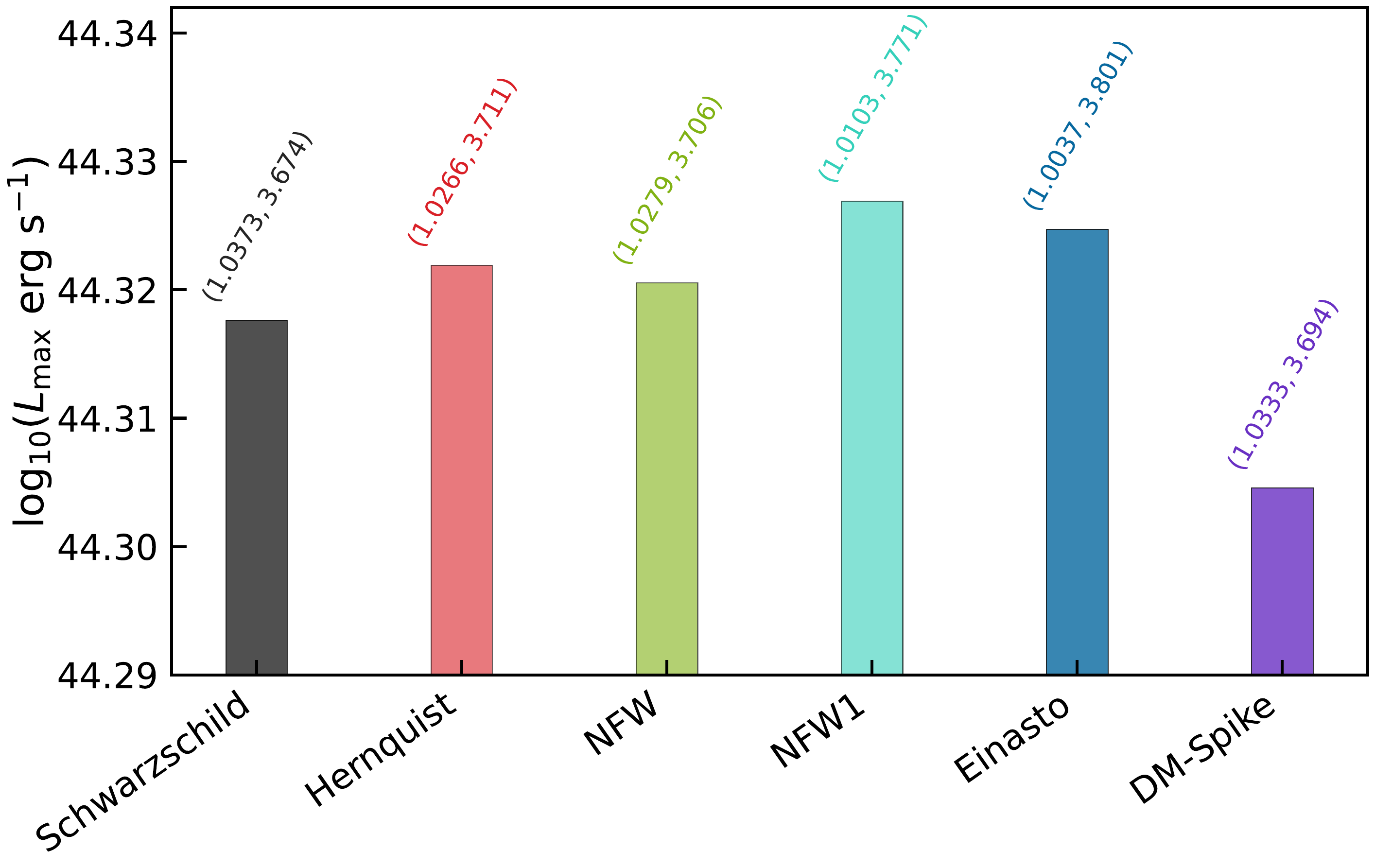}
    \caption{Maximum bolometric luminosity for different dark matter profiles. Here, we choose $M_{\rm halo}=10^2 M_{\rm BH}$ and $a_0=10^4 M_{\rm BH}$. See the text for details.}

    \label{fig:maxlumi} 
\end{figure*}

\section{Summary and Discussion} \label{sec:concl}

In this work, we study the advection-dominated transonic hot accretion flow of an inviscid relativistic fluid onto a galactic SMBH described by the metric (\cref{eq:metric}). We model the geometry of the galactic BH as a spherically symmetric central Schwarzschild BH immersed in a DM halo with radial density distributions. We represent the DM halo as an ideal gas composed of cold collisionless DM particles. In the absence of hierarchical mergers~\cite{Merritt:2002vj}, our assumptions imply a seed BH at the centre of a galaxy, dressed with the relativistic distribution of DM~\cite{Sadeghian:2013laa} and a density spike in the inner region~\cite{Vasiliev:2007vh, Shapiro:2016ypb, Ullio:2001fb, Merritt:2003qk, Gnedin:2003rj, Bertone:2005hw}.  Following~\cite{Speeney:2024mas}, we have modelled this DM density spike in two different ways: (i) Hernquist, NFW, and Einasto distributions, truncated at $r=4M_{\rm BH}$ and (ii) a fully relativistic HQ-type DM spike model where we assume the HQ-type DM spike grows adiabatically from an initial Hernquist type distribution. We explicitly assume the central BH to grow adiabatically~(\cref{fig:density-prof}). To obtain the effective spacetime metric in the presence of DM, we use the Einstein cluster formalism. This leads to the anisotropic nature (zero radial pressure) of the DM~\cite{Einstein:1939ms,Geralico:2012jt}. Thus, within the timescale of baryonic accretion, our analysis assumes the accretion rate of the DM particles being negligible. Furthermore, the analysis also ensures that the stress-energy tensor remains covariantly conserved and all the energy conditions are satisfied.

We maintain a strict hierarchy of the different length scales involved in the problem, $M_{\rm BH}\leq M_{{\rm halo}}\leq a_0$ and $r_{\rm out}\leq a_0$, $r_{\rm out}$ being the outer edge of the accretion disk.

Although the presence of the DM halo does not change the location of the event horizon ($r_{\rm h}=2 M_{\rm BH}$), it greatly modifies the geometry of the surrounding spacetime (including the location of the lightrings and ISCO).
This change in geometry, in turn, reduces the effective gravitational potential experienced by an infalling fluid element (\cref{fig:eff-pot}). The Keplerian angular momentum also shows a significant deviation from that of Schwarzschild BH, allowing flow with higher specific angular momenta to reach the event horizon (\cref{fig:kep}). We observe that in all the cases, the presence of DM halo shifts the critical points to smaller radii, affecting the solution topologies (\cref{fig:e_lm,fig:Mh-a0,fig:Solution}). The primary observable in our study is the luminosity of the emitted Bremsstrahlung radiation. 
Since the bolometric luminosity for I-type accretion flow is known to be the highest~\cite{Patra:2023epx,Sen:2024rna}, we chose the flow parameters such that the flow passes only through the inner saddle type critical point.

We observe (\cref{fig:solprop}) that the gravitational effect of the DM halo weakens the inflow of the accreting fluid, resulting in reduced flow velocity. The gravitational effect of the DM halo also increases the local flow temperature compared to that of a Schwarzschild BH.  Since the mass accretion rate is assumed to remain constant throughout the accretion flow, the local mass density of the accreting fluid is always inversely proportional to the local velocity of the flow. Again, since the specific energy of the flow remains constant, the lower flow velocity implies a lower kinetic energy and higher thermal energy of the accreting fluid, resulting in an increased local temperature (particularly, close to the inner edge of the disk). The increase in local flow temperature results in an increased thermal pressure, which increases the relative disk height, making the accretion disk quasi-spherical. Though the optical depth ($\tau$) of the disk still remains small, the increased disk height reduces the effective portion of the disk, contributing to the disk luminosity. 
This is more prominent for highly compact halos with a higher halo mass. The SED also shows a significant deviation from that of the vacuum Schwarzschild BH (\cref{fig:sed}). Since the dominant contribution to the luminosity for I-type accretion comes from the inner portion of the accretion disk, where deviation from the vacuum Schwarzschild geometry is strongly governed by the presence and nature of DM spike (overdensity), the variation in luminosity for the same set of halo and flow parameters indeed encode information about the nature of the density spike. 
The variations of the luminosity of the emitted radiation and properties of the accretion flow increase with the mass and compactness of the DM halo (\cref{fig:plumi}). However, in the case of the HQ-type DM spike distribution, the nonlinear dependence of the density profile on the halo mass changes the flow properties significantly with the halo mass (see \cref{app:DMspk} for details). 

Our results indicate that SMBHs in distant galaxies can, in fact, be less massive than expected if the DM contributions are considered. Since there exist tight correlations between the mass of the SMBHs and the properties of the host galaxies (such as the bulge mass~\cite{1998AJ....115.2285M, 2003ApJ...589L..21M} and velocity dispersion~\cite{2003ApJ...589L..21M, 2002ApJ...574..740T, 2003ApJ...589L..21M}) leading to coevolution of the SMBHs and the host galaxies our results suggest modifications in our understanding of galactic evolution~\cite{Kormendy:2013dxa}. 

 Indeed, the presence of DM and the nature of its distribution around the SMBH strongly influence the geometry of the surrounding spacetime. However,  the variation of bolometric luminosity may not be significant to infer the exact nature of the DM spike close to the BH. Thus, further research is required on more accurate modelling of the DM density spike around the SMBH and its correlation with a more realistic accretion model. 
In this regard, we mention the limitations of the accretion model used in our work. As a first step to address the present problem involving multiple length and time scales, we assume the accretion flow to be governed by ideal fluid dynamics, free from any dissipation mechanism. This results in a constant angular momentum of the fluid throughout the accretion flow as viscosity is ignored. Note that in the inner regions of the disc, the viscous timescale often exceeds the infall timescale of the accreting matter. As a result, the flow does not have sufficient time to redistribute angular momentum outward through viscous interactions, rendering the inner accretion effectively inviscid~\cite{Fukue-1987,Chakrabarti-1989}(see also Appendix A of \cite{Dihingia:2019xdx}).

Incorporating other dissipation mechanisms, such as radiative cooling and thermal conduction, are expected to regulate the disk temperature, influencing the spectral emission properties. Furthermore, throughout our analysis, we consider the electron and ion temperatures to be related by a simple scaling factor. We also disregard the presence of structured magnetic fields in this work.
 
The present findings obtained using these considerations are expected to be in agreement at least qualitatively with the observationally inferred systems such as Sgr A* or M87*, where magnetic fields, turbulance, viscosity, and multi-temperature plasma dynamics play central roles. It is worth mentioning that the implementation of these issues is beyond the scope of the present work, and we intend to take this up in our future endeavor.

Moreover, the predicted enhancements in disk luminosity due to the presence of a dense dark matter spike could be degenerate with other astrophysical factors: black hole spin alters the innermost stable circular orbit and thereby the radiative efficiency~\cite{Shapiro1974, McKinney-Gammie2004, Li-etal2004}; magnetic fields can significantly influence angular momentum transport and energy dissipation (e.g., via magneto-rotational instability)~\cite{Lynden-Bell1969, Shapiro1973, Chakrabarti-Mandal2006, Sarkar-Das2018}; and time variable accretion rates can mimic or obscure dark matter induced signatures~\cite{Shakura-Sunyaev1973, 1980A&A....88...23P, Chakrabarti-Titarchuk1995}. These degeneracies are particularly pertinent in the context of the EHT, which probes the innermost regions of accretion flows, as well as in indirect dark matter detection efforts, where excess emission may be attributed to non-baryonic processes. Future extensions incorporating spin (via Kerr or Kerr-like metrics) and magnetohydrodynamic (MHD) effects are essential to disentangle dark matter contributions from baryonic degeneracies and to more robustly predict observables that can be confronted with high-resolution EHT data.

\section{Acknowledgment}

Authors thank the anonymous reviewer for useful suggestions and comments that help to improve the quality of the paper. The work of AC is supported by the National Postdoctoral Fellowship of the Science and Engineering Research Board (SERB), ANRF, Govt. of India (File No.: PDF/2023/000550). SC acknowledges support of MATRICS research grant awarded by Science and Engineering Research Board (SERB), ANRF, Govt. of India through grant no. MTR/2022/000318. 

\appendix
\section{Keplerian angular momentum}\label{app:kep}
One of the key conditions for transonic hot accretion flow is that the constant flow angular momentum must be lower than the Keplerian angular momentum ($\lambda_{K}$).  
The Keplerian angular momentum is obtained by the vanishing gradient of the effective potential
\begin{equation}
    \frac{d\Phi^{\rm eff}_e}{d r}=0~.
\end{equation}
This implies that matter rotating in a Keplerian orbit will stay in equilibrium as the centrifugal force balances the effective gravitational attraction. The variation of the Keplerian angular momentum for the different DM profiles is shown in \cref{fig:kep}. We note that the presence of the DM halo always increases the Keplerian angular momentum from that of Schwarzschild BH. This implies that advection-dominated transonic hot accretion flow in the presence of DM halo can have higher specific angular momentum. Except for highly compact halos, the Einasto distribution shows the highest Keplerian angular momentum at any radius. 

The minima of the Keplerian angular momenta correspond to the radius of the marginally stable orbit~\cite{Abramowicz:2011xu, Rezzolla:2013dea}. We observe that the radius of the marginally stable orbit (ISCO, see \cref{eq:r-isco}) shifts inwards due to the presence of the DM halo, signalling a weaker inflow. This is consistent with the observation in~\cite{Heydari-Fard:2024wgu}.
 \begin{figure}[htbp!]
        \includegraphics[width=\linewidth]{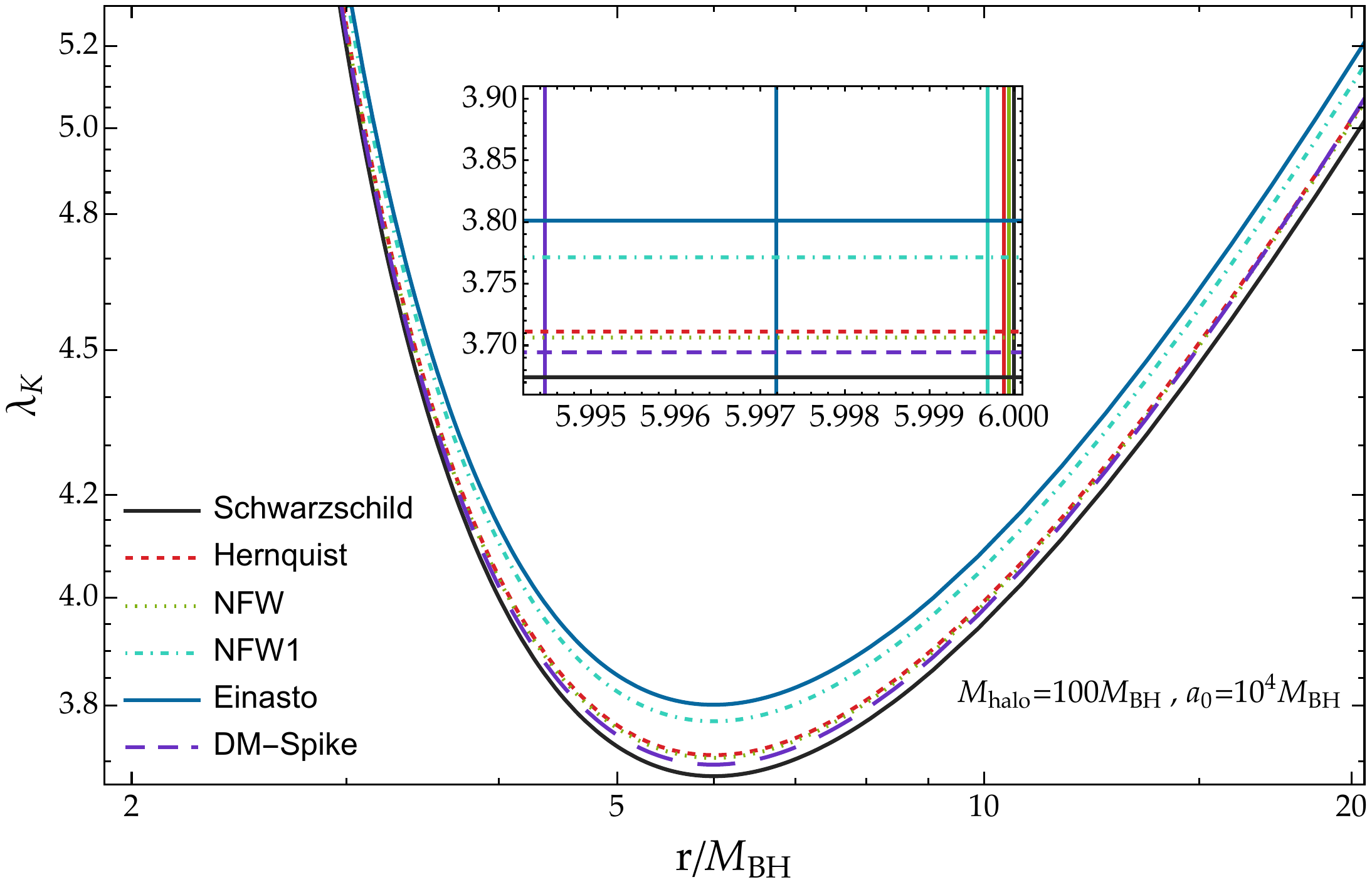}
        \caption{Variation of the Keplerian angular momentum ($\lambda_{K}$) with the radial coordinate ($r$) in the unit of the BH mass for different DM halo profiles. The vertical lines in the inset show the location of the marginally stable orbit (minima of $\lambda_{K}$) of the corresponding DM halo distribution.}
        \label{fig:kep}
    \end{figure}
    
 
\section{Hernquist type DM spike}\label{app:DMspk}
In this appendix, we plot the nonlinear dependence of the density profile of the HQ type DM spike with the halo mass, the associated effective potential~ \cref{eq:eff-pot} in \cref{fig:DMSpk-dens-m}. The resulting changes in the location and temperature at the critical point for I-type accretion flow with $\mathcal{E}=1.001$ and $\lambda=3.674$ is shown in \cref{fig:DMSpk-rc-Tc}. Figure~\ref{fig:DMSpk-rc-Tc} also shows the radius $r_0$ at which the aspect ratio becomes unity. Note that the maximum value of $r_0$ is set at 1000$M_{\rm BH}$ for the current analysis. Other flow properties strongly depend on the location and temperature of the critical point. This explains the nonlinear behaviour of the flow properties and luminosity of the HQ type DM distribution shown in \cref{sec:solprop} and \cref{sec:lumi}.

\begin{figure}
\includegraphics[width=0.45\textwidth]{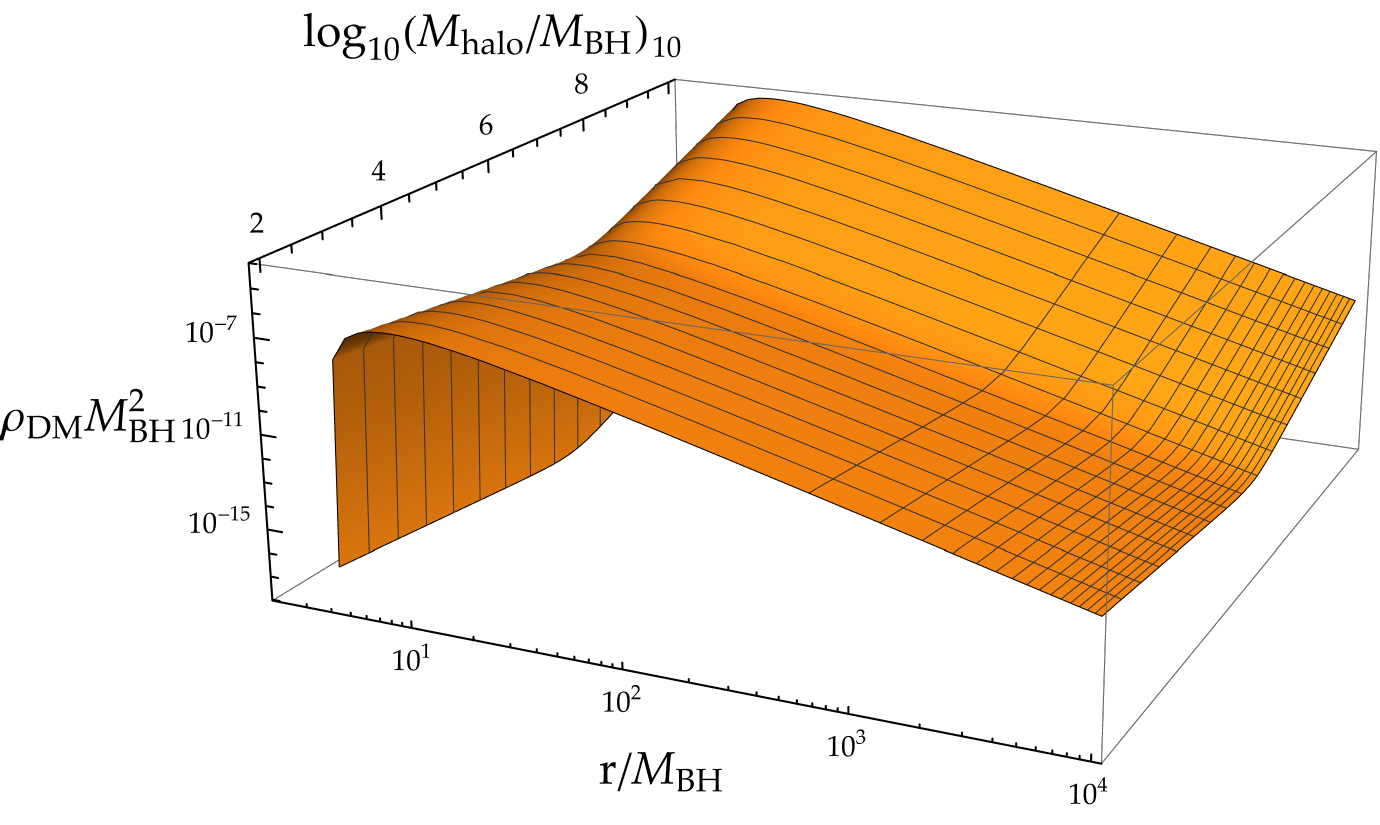}
\includegraphics[width=0.45
\textwidth]{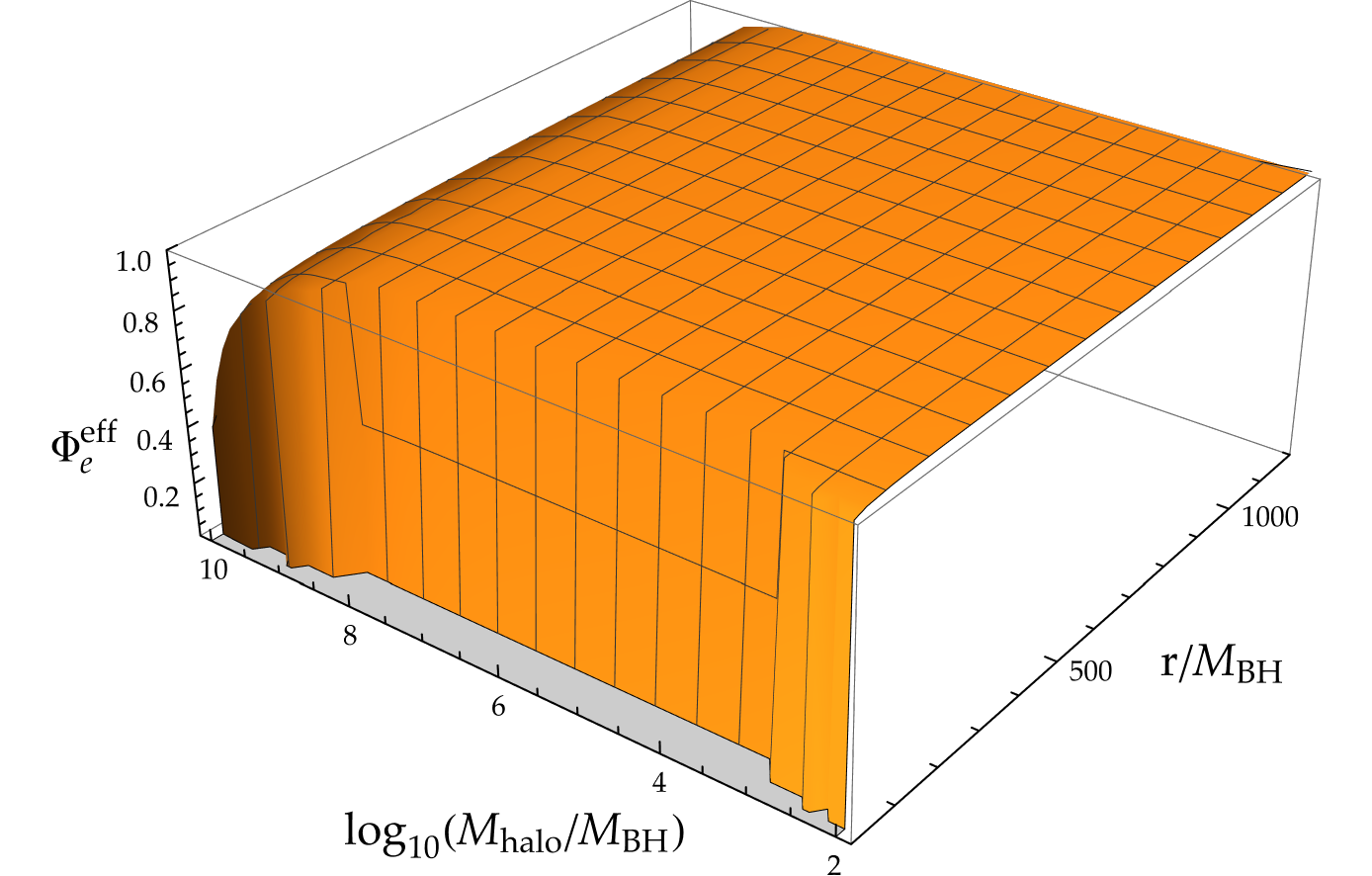}
    \caption{Variation of the halo density and effective potential of HQ type DM spike profile with the halo mass for a fixed DM halo compactness of 0.01.}
    \label{fig:DMSpk-dens-m} 
\end{figure}
\begin{figure}
    \includegraphics[width=0.45\textwidth]{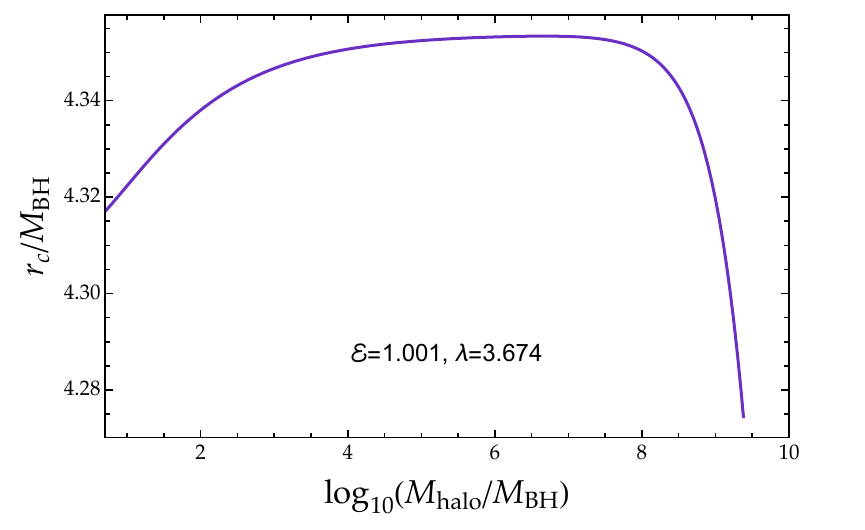}
\includegraphics[width=0.45\textwidth]{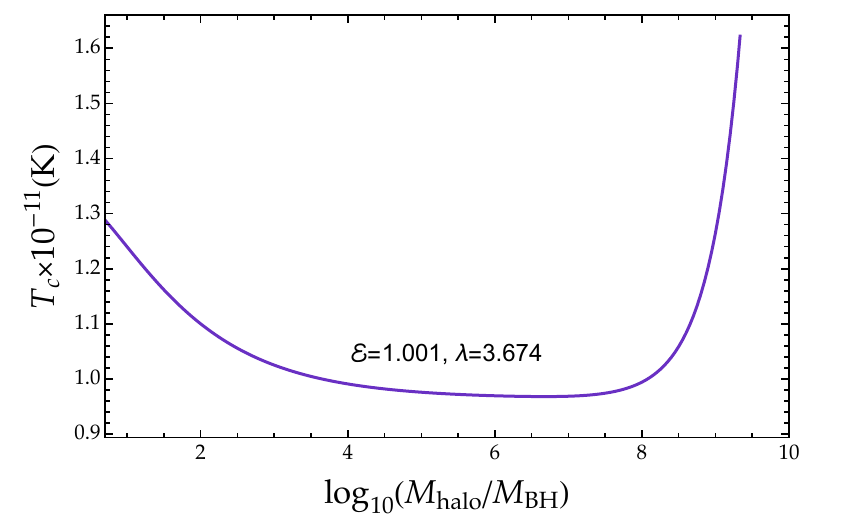}
\includegraphics[width=0.45
\textwidth]{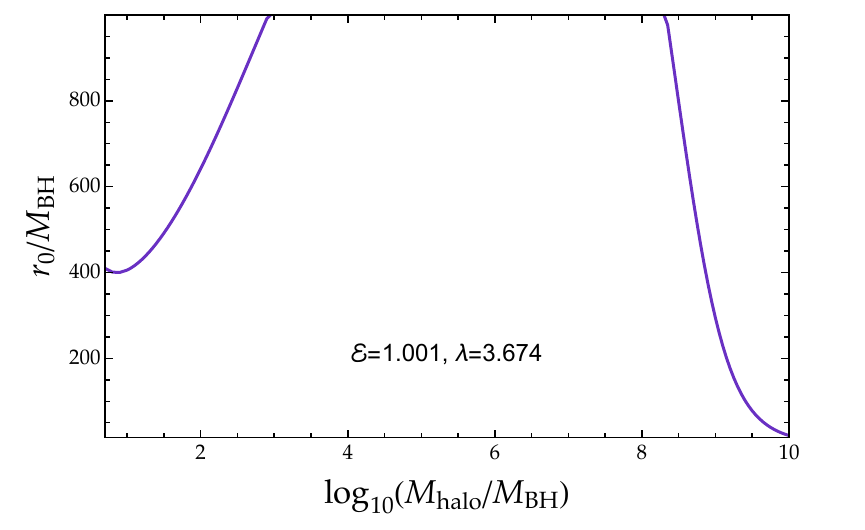}
    \caption{Plot of the critical points ($r_c$) (top) and temperature ($T_c$) of the critical point (middle) for a Hernquist type DM halo with $M_{\rm halo}$ around a galactic SMBH with $\mathcal{E}=1.001$ and $\lambda=3.674$.  In the bottom panel, variation of radial distance $r_0$ at which the relative disk height ($H/r$) becomes unity, is shown. Here, the maximum value of $r_0$ is set at 1000 $M_{\rm BH}$.
    }
    \label{fig:DMSpk-rc-Tc}
\end{figure}

\end{balance}

\newpage

\bibliography{ref}

\end{document}